This manuscript was accepted by Phys. Plasmas. Click here to see the version of record.

# Impact of ELM control techniques on tungsten sputtering in the DIII-D divertor and extrapolations to ITER

T. Abrams[a], E.A. Unterberg[b], D.L. Rudakov[c], A.W. Leonard[a], O. Schmitz[d], D. Shiraki[b], L.R. Baylor[b], P.C. Stangeby[e], D.M. Thomas[a], H.Q. Wang[a]

[a]*General Atomics, San Diego, CA 92186-5608, USA*
[b]*Oak Ridge National Laboratory, Oak Ridge, TN 37831, USA*
[c]*University of California San Diego, La Jolla, CA 92093-0417, USA*
[d]*University of Wisconsin Madison, Madison, WI 53706, USA*
[e]*University of Toronto Institute for Aerospace Studies, Toronto, M3H 5T6, Canada*

**Abstract**

The free-streaming plus recycling model (FSRM) has recently been developed to understand and predict tungsten gross erosion rates from the divertor during edge localized modes (ELMs). In this work, the FSRM was tested against experimental measurements of W sputtering during ELMs, conducted via fast WI spectroscopy. Good agreement is observed using a variety of controlling techniques, including gas puffing, neutral beam heating, and plasma shaping to modify the pedestal stability boundary and thus the ELM behavior. ELM mitigation by pellet pacing was observed to strongly reduce W sputtering by flushing C impurities from the pedestal and reducing the divertor target electron temperature. No reduction of W sputtering was observed during the application of resonant magnetic perturbations (RMPs), in contrast to the prediction of the FSRM. Potential sources of this discrepancy are discussed. Finally, the framework of the FSRM is utilized to predict intra-ELM W sputtering rates in ITER. It is concluded that W erosion during ELMs in ITER will be caused mainly by free-streaming fuel ions, but free-streaming seeded impurities (N or Ne) may increase the erosion rate significantly if present in the pedestal at even the 1% level. Impurity recycling is not expected to cause significant W erosion in ITER due to the very low target electron temperature.

## 1. Introduction

The tokamak instability known as the edge localized mode (ELM) is a periodic expulsion of plasma that occurs during a variety of H-mode scenarios [1]. The ELM instability originates primarily within the edge of the confined region known as the pedestal, and results in intense heat and particle pulses to the divertor targets within a very short time ($\lesssim$1 ms) after the ELM onset. The suppression or mitigation of the ELM heat and particle flux to the divertor target regions represents one of the most important ongoing plasma-materials interactions (PMI) challenges for ITER and future devices [2]. The limits on ELM size in ITER are typically parametrized by the total ELM energy fluence to the divertor target in units of (MJ $m^{-2}$) [3, 4] or of the "damage parameter" with units MJ $m^{-2}$ $s^{1/2}$ [2] in order to prevent melting or cracking of the tungsten divertor tiles, particularly on toroidal or poloidal leading edges of the W lamella [5].

In this work a slightly different limit on the ELM size is considered- the total W erosion of the divertor surface resulting from physical sputtering due to the incoming ELM ion flux. Erosion of the tungsten divertor targets can lead to high-Z contamination of the core plasma, causing strong radiative cooling and core fuel dilution [6, 7], thereby decreasing fusion performance. In addition, if the duty cycle of the device is sufficiently high, erosion of the divertor surfaces may determine the overall lifetime of these components before replacement. ITER will operate with a partially detached divertor with relatively cold electron temperatures (< 5 eV) at the strike point to mitigate W sputtering [8]. The electron temperature may increase moving radially away from the divertor targets, but this region is expected to be coated with beryllium deposits, shielding the W surface from erosion [9].

During unmitigated ELMs, however, the physical sputtering of tungsten from the divertor target region may still be substantial in the detached, near-target region due to ELM "burn-through." This effect has been observed in recent experiments in the JET ITER-like wall (ILW) [10]. This motivates developing a validated understanding of the physical mechanisms which drive the gross and net erosion of tungsten divertor components during ELMs. Recently the free-streaming plus recycling model (FSRM) [11] was proposed as a relatively simple 1D model to account for W gross erosion during ELMs, but only limited validation of the FSRM predictions for tungsten gross erosion has been performed. Recent validation studies of the FSM in the JET-ILW divertor have been reasonably successful [12, 13], but the crucial effect of particle recycling has not yet been incorporated in those models. The present work extends the validation of the FSRM calculations of W sputtering to a variety of pedestal and ELM regimes, with extensive variation in injected power and magnetic shaping to vary the ELM size and ELM frequency. In addition, the effect of two common ELM mitigation techniques, pellet pacing and resonant magnetic perturbations (RMPs), on the overall W sputtering rate in the DIII-D divertor is investigated in the framework of the FSRM. Finally, this model is utilized to extrapolate the gross erosion rate of tungsten during unmitigated ELMs on ITER, including the effect of impurity seeding, which will be essential in ITER to reduce the heat flux to the divertor targets but may substantially increase W sputtering during ELMs.

## 2. The Free-Streaming plus Recycling Model for ELMs

The Free-Streaming plus Recycling Model [11] is an extension of the popular Fundamenski-Moulton Free-Streaming Model [14, 15], FSM, an analytic model describing ELM heat and particle fluxes. The general conception of this model is shown in Figure 1, where magnetic field lines have been 'unwrapped' into a two-dimensional plane. The FSM assumes that a Gaussian-distributed plasma filament is transported out of the plasma core and streams rapidly along magnetic field lines in the scrape-off-layer (SOL) to the divertor targets. The inputs to the FSM are the density and temperature at the pedestal top, $n_{e,ped}$ and $T_{e,ped}$, the parallel length of the ELM filament, $L_{ELM}$, and the parallel intra-ELM magnetic connection length, $L_{\parallel}$. The FSM assumes the ELM filament does not interact with the background plasma and that sheath effects are unimportant. Expressions can be derived for resulting electron density, ion flux, and heat flux to the divertor targets, given by $n_{e,FS}(t)$, $\Gamma_{\parallel,FS}(t)$, and $q_{\parallel,FS}(t)$, respectively. The assumption $L_{ELM} = 2\pi R q_{edge}$ is made- that the parallel extent of the ELM filament is equal to the inter-ELM plasma connection length, which produces reasonable agreement with experimental measurements of the ELM energy fluence on multiple devices [3, 16]. The further assumption $L_{\parallel} = 9L_{ELM}$ is consistent with nonlinear simulations [17] as well as experiments on JET-ILW [12] and DIII-D [11], and will be used in the present work.

The impact of particle recycling was recently incorporated into the FSM by introducing an effective particle recycling coefficient, $R_{\text{eff}}$, creating the free-streaming plus recycling model, FSRM [11]. Recent experimental observations on DIII-D [11, 18] have inferred a value of $R_{\text{eff}}$ in the divertor of approximately 0.96. Recycling is included in the FSRM by assuming that energetic ions from the pedestal ions striking the divertor surface are recycled as neutrals with probability $R_{\text{eff}}$, converted to ions via electron-impact ionization, thermalized with the inter-ELM divertor background plasma, and finally return to the target. The FSRM predictions of divertor ion fluence were benchmarked against an extensive DIII-D database that spanned a large range of neutral beam injection (NBI) power and pedestal/divertor conditions [11]. Good agreement was observed between the DIII-D database and FSRM predictions, implying that the FSRM is sufficiently accurate to replicate the dynamics of the ELMy particle flux. Similar observations were made for the ELM energy fluence.

The FSRM was then extended to make predictions of the time-dependent divertor erosion rate. For tungsten plasma-facing components (PFCs) in DIII-D, this includes physical sputtering caused by free-streaming D and C ions originating from the pedestal top, and by D and C ions recycling from the divertor surface (Equation (5) in [11]). Physical sputtering yields $Y$ are calculated with the SDTrim.SP code using a 45° ion impact angle. The magnetic field angle was generally between 1.5° and 2.5° in these studies. However, the impact angle of impinging ions is substantially more normal due to sheath and gyro-orbit effects. Calculations of the ion impact angle during ELMS near the JET-ILW target have shown a broad distribution between 0° to 85° from surface normal [19]. However, as shown in Figure 2, the D→W and C→W sputtering yields at 45° are approximately the average values between 0° and 85° for the relevant impact energies. Thus a 45° impact angle is considered a reasonable proxy for the effective sputtering yield of tungsten for both impinging species. Similarly, as the recycling species are assumed to be thermalized with the background plasma, an angular distribution peaked between 45 degrees and 65 degrees from surface normal is expected [19, 20, 21]. Given that surface roughness effects will shift the impact angle distribution more in the normal direction [21], a representative value of 45 degrees for the impact angle is considered reasonable.

W→W self-sputtering is neglected in the FSRM. Due to short ionization mean free path of neutral W relative to the gyro-radius of $W^+$, a large fraction of sputtered W is expected to be promptly re-deposited. Using the empirical formula provided by Equation (2) of [22] results in a prompt re-deposition fraction of 0.96 for typical intra-ELM divertor plasma conditions ($n_{e,div}$=5x$10^{19}$ $m^{-3}$, $T_{e,div}$=30 eV). Entrainment of $W^+$ with either the free-streaming plasma or the background plasma is not expected because the $W^+$ gyro-period is much smaller than collisional time scales. If the neutral W is ionized entirely outside the sheath, this implies ion impact energies of ~100 eV, which corresponds to a W→W sputtering yield of ~0.1. This effectively results in a $0.1+0.1^2+0.1^3$= 11.1% increase in W sputtering. In practice this is an upper bound on the self-sputtering contribution because some ionization will occur within the sheath, leading to lower average W ion impact energy. Thus self-sputtering is not expected to be a significant effect.

The impact energy of the free-streaming ions is simply equal to the parallel heat flux divided by the parallel ion flux, Equations (2) and (3) in [11]. A Maxwellian energy distribution plus sheath potential $3nT_{e,div}$ is specified for the recycling ions because the FSRM assumes that the recycling species are thermalized with the background plasma. The $C^{6+}$ impurity at the pedestal top is obtained from measurements using DIII-D edge charge exchange recombination (CER) spectroscopy system [23]. The time constant for $C^{6+}$ impurities transported from pedestal top to the divertor surface is significantly smaller than the $C^{6+}$→$C^{5+}$ recombination time [24], and thus recombination for the free-streaming impurities is ignored. The recycled C ion impurity flux is assumed to be all $C^{2+}$ ions that have recombined from $C^{6+}$, which is approximately the average charge state predicted by modeling for the background plasma [25, 20]. The stoichiometric fraction of C (due to implantation) in the top W surface layers is assumed equal to 0.5, in line with the results of previous studies [24, 25], which effectively reduces the predicted W sputtering rate by a factor of 2.

Using the FSRM, examples of the time evolution of the divertor electron density, $n_{e,div}$, perpendicular ion flux, $J_{sat}$, perpendicular heat flux, $q_{\perp}$, and tungsten gross erosion rate, $\Gamma_W$, calculated by the FSRM are shown in Figure 3. The plasma parameters specified are fairly typical DIII-D H-mode attached divertor conditions ($n_{e,ped}$ = 4x10$^{19}$ m$^{-3}$, $T_{e,ped}$ = 600 eV, $T_{e,div}$= 25 eV, $\theta_{div}$ = 2.0°, $f_{C,ped}$ = 0.03) and an effective particle recycling coefficient $R_{\rm eff}$ = 0.96 is assumed. All traces indicate a fast rise time followed by an exponential-type decay phase, as is commonly observed for ELMs of various sizes [1]. The contributions to each parameter from the free-streaming component and recycling component of the ELM plasma, however, differ widely. The divertor electron density is almost entirely determined by the recycling ion flux, and thus by the value assumed for $R_{\rm eff}$. Both the free-streaming plasma and the recycling plasma contribute strongly to the ELMy ion flux. The intra-ELM heat flux is almost entirely caused by the free-streaming plasma filament because each recycling ion carries substantially less energy than the free-streaming ions. The W physical sputtering is also dominated by the free-streaming $D^+$ and $C^{6+}$ ions, with a small but non-zero contribution from the recycling $C^{2+}$ ions.

As an extension of this result, it is useful to understand how the FSRM predicts the contribution to the overall W erosion rate will change as a function of pedestal and divertor conditions. In Figure 4, the fractional contributions of free-streaming $D^+$ ions, free-streaming $C^{6+}$ ions, and recycling $C^{2+}$ ions are plotted against the typical range of pedestal temperatures achieved on DIII-D. The case of relatively low divertor temperature ($T_{e,div}$ = 10 eV) and high divertor temperature ($T_{e,div}$ = 30 eV) are included for comparison. The pedestal $C^{6+}$ impurity fraction is set to 2% in both cases. Note that recycling $D^+$ ions do not have sufficient impact energy to sputter W atoms in either of these cases and their contribution remains zero for all values of $T_{e,ped}$ and both values of $T_{e,div}$ investigated.

For the high $T_{e,div}$ case, all three W erosion pathways contribute roughly equally to the W sputtering rate at low pedestal temperature ($T_{e,ped}$= 300 eV). As $T_{e,ped}$ increases, the fractional contribution of the free-streaming $D^+$ ions to the overall gross W erosion rate increases, while the contribution of free-streaming $C^{6+}$ and recycling $C^{2+}$ ions decreases. This transition occurs because the physical sputtering yield of D on W increases strongly with ion impact energy (proportional to $T_{e,ped}$) in this regime, while the sputtering yield of C on W is relatively constant. For the low $T_{e,div}$ case, the contribution of recycling $C^{2+}$ ions to the W erosion rate is negligible for all pedestal temperatures. This is because a divertor electron temperature of 10 eV results in recycling $C^{2+}$ ions with impact energies very close to the threshold for C→W physical sputtering. The fractional contributions of the free-streaming species show similar trends to the high $T_{e,div}$ case, with free-streaming $C^{6+}$ being the most important contributor at low values of $T_{e,ped}$, but free-streaming $D^+$ becoming progressively more important as pedestal temperature increases. Note that in the FSRM formulation, all W erosion rates during ELMs are linearly proportional to pedestal density, and thus $n_{e,ped}$ does not affect the fractional contributions to the overall W erosion rates displayed in Figure 4.

## 3. Experimental apparatus and procedure

ELM-resolved measurements from a number of DIII-D edge and divertor diagnostics were leveraged to validate the calculations of the FSRM (Figure 5a). The results shown in this work involve two different experimental set-ups. The first, shown in Figure 5b, involved exposure of small (~20 cm$^2$) W-coated graphite samples of ~200 nm thickness via the use of the Divertor Materials Evaluation System (DiMES), a removable sample exposure probe in the DIII-D lower divertor [26]. The second, more comprehensive set of experiments was conducted on two toroidally symmetric W-coated tile arrays in the DIII-D lower divertor during a mini-campaign known as the Metal Rings Campaign (MRC) [27], as depicted in Figure 5c. The characteristic roughness of the W coatings was 0.25-0.5 μm as measured by coherence scanning interferometry. No substantial change in impact angle distribution is expected from this relatively smooth surface, but as discussed above, surface roughness will make the average ion impact angle somewhat more normal. This may result in a small over-prediction of the W physical sputtering yields (Figure 2).

In both cases, the primary diagnostic used for these studies was the DIII-D filterscope array [28]. The standard $D_\alpha$ filterscope channels were used for detection of ELM start times by the rising edge of the spectroscopic signal. Recently installed WI (400.9 nm) filterscopes [29] were used to infer the gross erosion rate of tungsten PFCs during ELMs via the S/XB method. This method assumes that the gross erosion rate from a material is proportional to the spectroscopic intensity of a neutral emission line times the ionization/photon coefficient, which is derived from atomic physics considerations [30]. For neutral tungsten influx, this process can be parameterized as $\Gamma_W(t) = \frac{S}{XB}(t)\int_0^\infty I_{WI}(t)dz$, where the W gross erosion rate is $\Gamma_W$, the neutral W line emission intensity is $I_{WI}$, and the $z$ direction is along the line of sight of the diagnostic (i.e., the WI filterscope). The WI S/XB coefficient is time-dependent because it is a function of divertor electron density, which tends to evolve substantially through the ELM cycle [11]. For example, for the case shown in Figure 3, assuming an inter-ELM divertor electron density of $10^{19}$ $m^{-3}$, the WI S/XB coefficient varies from 31 in the inter-ELM phase to 64 at the peak of the intra-ELM phase. The increase in the WI S/XB is roughly linear with the intra-ELM divertor density. Detailed plots of the time evolution of the WI S/XB coefficients for the three cases depicted in Figure 8 can be found in Figure 8 of [11]. The divertor electron density, $n_{e,div}$, and temperature, $T_{e,div}$, are measured by Langmuir probes and the Divertor Thomson Scattering (DTS) systems, as shown in Figure 5. The pedestal electron density $n_{e,ped}$ and electron temperature, $T_{e,ped}$, are diagnosed by the edge Thomson scattering array [31] and the pedestal $C^{6+}$ impurity density is measured by the edge CER system [23].

An overview of the experimental procedure used in this work is shown in Figure 6 for an H-mode discharge with an ELM frequency of approximately 40 Hz. First the discharge is established with reasonably constant line-average density, Figure 6a, with the outer strike point (OSP), Figure 6c, placed on the graphite tile radially outboard of the W-coated portion of the DIII-D lower divertor shelf. At 1900 ms the OSP is rapidly swept inward to a fixed position at the inner radius of the W-coated tile on the divertor shelf. The WI emission intensity from the divertor surface, monitored via the WI filterscope, Figure 6d, increases both during and between ELMs when the W-coated tiles are moved from the private flux region (PFR) into the common flux region (CFR), as expected. In this particular discharge, the ELM frequency begins to decrease around 2600 ms, resulting in increasing line-average density, increasing MHD stored energy (Figure 6c), as well as slightly larger ELMs. The time range of interest for this discharge is specified as 2100 to 2600 ms where relatively steady ELM conditions exist. A similar procedure is performed for each shot in the DIII-D database to determine the appropriate time range of interest.

The coherent averaging technique is used extensively in this work to increase the signal to noise ratio (SNR) from the WI emission signal, which tends to be rather weak due to the low sputtering yield of tungsten PFCs. The time base for each ELM event is shifted such that $t_{ELM}$ = 0 corresponds to the ELM start time. The traces are then overlaid as a function of time relative to the ELM start to check for consistency (Figure 7). Next the signals are binned into 0.1 ms averaging windows (relative to the 0.02 ms time resolution of the raw filterscope signal) and finally coherently averaged, increasing the SNR of the WI emission signal during ELMs. The standard deviation of the raw WI emission signals within each averaging period is considered as the error bar. This procedure is slightly different than some other works that use the temporal peak of the ELM signal as the common reference time, e.g., [3]. Given the amount of noise in the raw WI emission signals, however, it was determined that using the ELM start time as the common reference produced more reliable results for this study. Additionally, in certain cases close to the L-H power threshold, the ELM event induces a short (5 ms) H-L back-transition during the ELM decay phase, resulting in a brief period of intense W sputtering. In such cases the WI emission signal is extrapolated as an exponential decay based on the ELM behavior prior to the H-L back-transition.

## 4. Modification of W sputtering using ELM control/mitigation techniques

### 4.1 ELM control via pedestal fueling

An experiment was performed to study W sputtering during ELMs in the DiMES geometry (Figure 5b) in which varying amounts of $D_2$ gas were injected to modify the pedestal density and temperature [11]. The overall pedestal pressure remained relatively constant, consistent with peeling-ballooning models of pedestal stability for this parameter range, which is very similar to that explored in Figure 5 of [32]. The results of this experiment are summarized in Figure 8, in which the total number of W atoms sputtered per unit area for each case is plotted as a function of pedestal density. The lowest fueling case resulted in a relatively high temperature, low density pedestal with a strongly attached divertor. Similar to the solid lines in Figure 4, the FSRM predicts that most of the W sputtering is caused by the free-streaming $D^+$ and $C^{6+}$ ions in roughly equal proportion, with a smaller but non-negligible contribution (20%) of the tungsten gross erosion caused by the recycling $C^{2+}$ ions. The energy of the recycling $D^+$ ions is below the threshold for D→W physical sputtering and thus no W erosion is caused by the recycling main ions. Notably, the cumulative W gross erosion calculated by the FSRM is found to be consistent with the experimental measurement.

As the $D_2$ puffing rate is increased, a higher pedestal density is obtained at approximately the same pedestal pressure, resulting in a lower value of $T_{e,ped}$. This additional gas puffing also cools the divertor, resulting in a lower value of $T_{e,div}$. Higher pedestal density results in more free-streaming ion fluence to the target during ELMs, but the lower pedestal and divertor temperatures result in lower impact energies for the F-S and recycling ions, which lowers the W sputtering yields. These two effects approximately cancel out, resulting in nearly the same total W erosion per $m^2$ per ELM. This FSRM calculation also lies in quantitative agreement with the experimental measurement. Finally, at the highest $D_2$ puffing rate, the pedestal and divertor temperatures further decrease while $n_{e,ped}$ again increases. Here the decrease in ion impact energy, and corresponding lower C→W and D→W sputtering yields, dominates over the enhanced ion fluence to the targets, and the FSRM predicts that the W gross erosion rate decreases. This effect is also observed in the experiment, providing a robust validation of the FSRM.

### 4.2 ELM control via heating power and plasma shaping

As a further test of the Free-Streaming plus Recycling Model, an experiment was conducted during the DIII-D Metal Rings Campaign [27, 33] where the pedestal stability boundary and the ELM frequency were systematically modified by changing upper triangularity, $\delta_{up}$, and heating power, $P_{heat}$. These actuation methods provided the ability to modify the pedestal density and temperature without strongly impacting divertor conditions. Heating power and discharge fueling was provided by Neutral Beam

Injection (NBI), with no additional fueling introduced by gas puffing. The divertor plasma remained strongly attached ($T_{e,div} \geq 25$ eV) for all discharges in this experiment. This allowed for a more precise test of the FSRM- namely, how W sputtering changes due to modifications of the free-streaming ion population, with minimal changes to the sputtering yield of the recycling impurity ions.

An overview of the results of this experiment is shown in Figure 9. First a low-power, strongly shaped ($P_{NBI}$ = 3 MW, $\delta_{up}$ = 0.27) scenario was established with large, infrequent ELMs. This actuation regime produces a relatively high density, low temperature pedestal ($n_{e,ped}$ = 7.3x10$^{19}$ m$^{-3}$, $T_{e,ped}$ = 360 eV) with a strongly attached divertor ($T_{e,div}$ = 37 eV), a relatively high $C^{6+}$ impurity fraction in the pedestal ($f_{C,ped}$ = 3.9%), and very infrequent ELMs ($f_{ELM}$ = 9 Hz). The measured W sputtering rate through the ELM cycle is shown in Figure 9a with the calculations from the FSRM overlaid. The time-integrated values of the total W gross erosion per m$^2$ per ELM, $\Phi_W$, from the measurements and FSRM are also provided. Due to the relatively high C impurity fraction in the pedestal, in conjunction with the relatively low D→W physical sputtering yield associated with this pedestal temperature, the free-streaming $C^{6+}$ impurity ions provide the largest contribution to the W gross erosion rate. Both the free-streaming $D^+$ ions and recycling $C^{2+}$ ions, however, also cause noticeable W sputtering.

In the next scenario, Figure 9b, the strong shaping of the plasma boundary is eliminated while maintaining the same heating power. This change modifies the plasma stability boundary, resulting in higher frequency ELMs ($f_{ELM}$ = 26 Hz), a lower pedestal density ($n_{e,ped}$ = 5.8x10$^{19}$ m$^{-3}$), but similar electron temperature in the pedestal ($T_{e,ped}$ = 360 eV) and divertor ($T_{e,div}$ = 38 eV). The W gross erosion rate during ELMs is reduced because the lower pedestal density results in lower free-streaming and recycling ion fluence to the target, with approximately the same physical sputtering yields because $T_{e,ped}$ does not change. This same trend is reproduced by the FSRM, which indicates that the relative balance of the W sputtering mechanisms remains about the same but the magnitude of each sputtering source decreases by about 15%

Finally, in the third scenario, Figure 9c, this weak shaping is maintained while increasing the NBI power, resulting in a lower pedestal density ($n_{e,ped}$ = 4.2x10$^{19}$ m$^{-3}$) and higher temperature ($T_{e,ped}$ = 500 eV), while again maintaining similar divertor conditions ($T_{e,div}$ = 29 eV). The W gross erosion rate is reduced by the lower free-streaming and recycling ion fluence to the target, but increased by the higher W physical sputtering yields caused by the higher value for $T_{e,ped}$. Because the total W gross erosion per ELM decreases from case (b), the decrease in $n_{e,ped}$ is clearly the dominant effect, and this decrease is also captured by the FSRM. A larger fraction of the W sputtering is caused by the free-streaming $D^+$ main ions relative to the low-power, cold pedestal case. It is also noted that in cases (b) and (c) the W gross erosion rate after the ELM actually dips below the value before the ELM onset. This is attributed to the "cold pulse" phenomenon, resulting in a decrease in $T_{e,div}$ near the end of the ELM phase, which has been examined in other works [11, 34]. The FSRM does not include the physics of the cold-pulse effect, which may explain why the FSRM slightly over-estimates the W sputtering rate for $t - t_{ELM}$ > 2 ms.

In transitioning from scenario (a) to (b) to (c), the intra-ELM W gross erosion decreases only slightly while the ELM frequency rises substantially. Therefore the total W source due to ELMs increases at high ELM frequency. This observation is consistent with previous studies on JET-ILW conducted via heating power scans [35]. Such results do not necessary imply, however, that low-frequency ELM regimes are favorable to reduce W core contamination due to physical sputtering by ELMs. Higher ELM frequencies are generally more efficient in "flushing" high-Z impurities from the edge of the confined region [36, 37].

large, low frequency ELMs may also be more efficient at penetrating through the divertor detachment front in devices such as ITER [10]. Further consideration of the balance between ELM size and ELM frequency in the context of W sputtering from the divertor can be found in Section 5.

In Figure 10, the experimentally measured W gross erosion per ELM, per unit area, is compared to the calculations of the FSRM for the entire W-DiMES and MRC experimental database, including during the use of ELM mitigation techniques (Section 4.3, below). The predictions for the ITER half-field case and full-field case, discussed in Section 5, are also overlaid. Good agreement is observed between the predictions of the FSRM W sputtering model and the experimental data over a broad parameter range. This provides further confidence that the physics model developed for the FSRM provides a reasonably accurate physical picture of W physical sputtering during ELMs in DIII-D.

By analyzing the residuals, however, it is clear that the intra-ELM W erosion measurements are not evenly distributed above and below the FSRM predictions. The measurements tend to be over-predicted by the model, in several cases up to a factor of 50%, as shown by the bottom dashed line in Figure 10. One potential contributing factor is surface roughness effects. As discussed above, surface roughness tends to shift the ion impact angle distribution in the normal direction, which will result in some reduction of the W physical sputtering yields. Given that the surface roughness of these coatings is relatively low, and $Y_{C\rightarrow W}$ and $Y_{D\rightarrow W}$ vary relatively slowly with impact angle, this effect cannot fully account for the over-prediction of the FSRM.

Since the FSRM is a 1D model, it should be verified that this validation exercise is not strongly impacted by radial variations of the intra-ELM W gross erosion rate. The radial footprint during ELMs tends to broaden substantially relative to the inter-ELM phase [38, 39], but if the W-coated tile is too far from the OSP, the W gross erosion rate may be reduced relative to the value at the OSP, causing an over-prediction by the model. The FSRM calculations for each discharge normalized to the experimental measurements are plotted in Figure 11a as a function of the radial distance between the center of the WI filterscope view, $R_{fs}$, and the OSP. No correlation is observed between the model vs. measurement discrepancies and radial distance of the filterscope view from the OSP for values of $R_{fs}$-$R_{OSP}$ < 4 cm. This indicates that the W sputtering profile during ELMs is relatively broad and not sensitive to the measurement location. For values of $R_{fs}$-$R_{OSP}$ > 5 cm, however, the FSRM begins to over-predict the experimental measurements, indicating that a radial decay of the intra-ELM W gross erosion occurs sufficiently far from the strike points, as expected. Therefore, the radial distance between the center of the WI filterscope chord and the OSP is used as a discrimination tool to determine which discharges are included in the DIII-D ELMy W sputtering database. Any cases with $R_{fs}$-$R_{OSP}$ > 5 cm are excluded and are not shown in Figure 10.

Additionally, it would also not be surprising if a simple fixed C/W mixed-material fraction of 0.5 on the W plasma-facing surface were entirely sufficient to account for the full range of PMI physics effects occurring through the ELM cycle. An over-prediction of the W gross erosion rate would imply that the actual C content on the W surface may be higher than the value assumed in the model, or that substantial temporal variation of the C/W mixed-material fraction occurs. Previous ERO simulations on DIII-D [25] have indicated that, in steady state, the C/W surface ratio tends to be proportional to the C ion flux fraction to the target. To test this hypothesis for the ELMy regime, the ratio of the FSRM calculations to the experimental measurements are plotted as a function of the $C^{6+}$ content at the pedestal top in Figure 11b. Again, no clear correlation emerges between the pedestal C content and model/measurement discrepancy when the W-coated PFCs are in close proximity to the OSP. Interestingly, however, the model/measurement ratio does tend to increase with $f_{C,ped}$ farther from the OSP, which, in steady state, is believed to be a region of net C deposition [40, 41]. Finally, given that almost no data points lie below the 1:1 line in Figure 11, a C/W mixed-material fraction of 0.5 appears to be the typical lower bound achievable for DIII-D H-mode plasma exposures of tungsten PFCs.

### 4.3 ELM mitigation techniques

#### 4.3.1 Pellet Pacing

As a further test of the FSRM and to explore ELM regimes relevant to ITER and next step devices, additional experiments were conducted to investigate the impact of ELM mitigation techniques on the ELMy W sputtering rate. The use of $D_2$ pellet pacing to mitigate ELMs has been previously demonstrated in a range of DIII-D plasma scenarios [42]. In this experiment, $D_2$ pellets were launched from near the low-field side midplane to trigger ELMs. First a baseline case was established with large, low frequency ELMs ($f_{ELM}$ = 10 Hz, $\Delta W_{MHD}$ = 81 kJ) with a relatively high-density, low temperature pedestal. Next, the discharge was repeated while injecting $D_2$ pellets at a frequency of 60 Hz. Pellet pacing decreased the ELM size to 40 kJ while increasing the ELM frequency to 20 Hz. The total power exhausted by ELMs remained approximately constant at about 20% of the heating power, as is typically observed [43]. In the ELM-mitigated discharges, the pedestal density and temperature remained similar to the baseline scenario, but $f_{C,ped}$ decreased from 4.5% to 2.8%. Reduction of the pedestal C impurity content at increased ELM frequency is in line with previous studies [44]. $T_{e,div}$ also decreased from 32 eV to 14 eV due to pellet pacing, presumably due to gas build-up in the divertor caused by the addition of pellet fueling.

The effect of pellet pacing on the W sputtering rate during ELMs in shown in Figure 12. The no-pellet scenario, Figure 12a, shows very similar pedestal and divertor conditions to the discharge studied in Figure 9a. The total W sputtering per ELM is relatively high and the strongest sputtering mechanism is free-streaming $C^{6+}$ impurity ions due to the high pedestal C impurity content and significantly higher C→W physical sputtering yield, relative to D→W, at this low value of $T_{e,ped}$. After the application of pellet pacing for ELM mitigation, the W sputtering rate decreases by about a factor of two. Because the ELM frequency also doubles, total W gross erosion *per second* due to ELMs remains approximately constant. As depicted in Figure 12b, the FSRM provides a fairly accurate depiction of the reduction of W sputtering via ELM mitigation. The free-streaming $D^+$ component of the W erosion does not change because the pedestal density and temperature remain constant. The lower pedestal C impurity content, however, results in a strong decrease in free-streaming $C^{6+}$→W sputtering, and the low value of $T_{e,div}$ essentially eliminates W erosion due to recycling $C^{2+}$→W. In contrast to the ELM control techniques above in which the W sputtering rate was modified primarily by changing the pedestal temperature and density, here it is demonstrated that the FSRM still reasonably accurately predicts changes to intra-ELM W sputtering while $n_{e,ped}$ and $T_{e,ped}$ remain constant.

#### 4.3.2 Resonant Magnetic Perturbations

The impact of Resonant Magnetic Perturbations (RMPs) on W sputtering rates during ELMs was also investigated during the MRC in conjunction with an experiment aimed at achieving ELM suppression. A low collisionality pedestal ($n_{e,ped}$ = 3.7x10$^{19}$ m$^{-3}$, $T_{e,ped}$ = 840 eV) was developed as the no-RMP reference case. These discharges were conducted in the so-called pumping configuration with the OSP placed on the floor array of W-coated tiles. In this experiment, a low-density, steady-state H-mode discharge at medium heating power ($P_{NBI}$ = 6.1 MW) was first established with the OSP placed on the W surface. Approximately 1000 ms later, the DIII-D internal coil set was energized to produce the RMP field with a toroidal periodicity $n = 3$ at a constant coil current of 5 kA. This actuation resulted in a strong reduction of $n_{e,ped}$, a commonly observed effect with the application of 3D fields in DIII-D at low collisionality [38], along with a 20% increase in $T_{e,div}$, while $T_{e,ped}$ remained nearly constant. Some evidence of ELM mitigation was observed via a decrease of the peak values of the on the $D_\alpha$ filterscope signals, but the values of $f_{ELM}$ and $\Delta W_{MHD}$ before and after application of the RMP remain approximately the same.

In Figure 13, the W gross erosion rate through the ELM cycle is displayed both before and after the application of the RMP. Note that the scale bar on the vertical axis is a factor of four larger here than in Figures 8 and 11, i.e., the W sputtering in this scenario is quite intense due to the high $T_{e,ped}$, leading to high physical sputtering yields for D→W and C→W. Before the RMP begins, the FSRM calculation agrees fairly well with the experimental measurement, although the decay time is somewhat over-predicted. After the application of the RMP, the intra-ELM W gross erosion rate actually remains approximately the same, or even increases slightly. This is in contrast to the trend predicted by FSRM, which suggests that the W sputtering should decrease after the application of the RMP due to the significantly reduced pedestal density and a slightly lower value of $T_{e,ped}$. This effect dominates over the small amount of additional W sputtering caused by the increase in $f_{C,ped}$ and $T_{e,div}$.

The FSRM under-predicts the peak value and over-predicts the decay time of the W gross erosion rate during application of the RMP. Relaxing the earlier assumption that $L_\parallel$ is fixed at a value of $9L_{ELM}$, the prediction of the FSRM is overlaid in Figure 13b assuming $L_\parallel = 6L_{ELM}$, indicating relatively good agreement with both the peak value and characteristic decay time of the measured W sputtering rate. As the cases depicted in Figure 10 represent the highest pedestal pressure and lowest collisionality in the ELMy W erosion database, this result may indicate that $L_\parallel$ may be affected by the relative position on the peeling-ballooning stability boundary. Previous JOREK simulations of a low-collisionality JET discharge [45] indicate that non-linear coupling between high-order toroidal mode numbers can result in faster decay times of the peak divertor heat flux (i.e., lower values of the effective $L_\parallel$) relative to when the high-order harmonics are excluded. The RMP may further modify the magnetic field topology in a manner that reduces the effective magnetic connection length between the ELM filament and the divertor target. A detailed MHD analysis of such effects is outside the scope of this paper but remains an important topic for future work.

## 5. Discussion and Implications for ITER

The FSRM can be used as a framework for W sputtering during ELMs to extrapolate what gross W erosion rates may be expected during unmitigated ELMs in ITER. In developing these extrapolations, the same assumptions for $L_{ELM}$ and $R_{\text{eff}}$ used in the DIII-D cases are applied to ITER. For the parallel intra-ELM magnetic connection length, $L_\parallel$, assumption $L_\parallel = 9L_{ELM}$ is used throughout, with the caveat that $L_\parallel/L_{ELM}$ may decrease somewhat for high-temperature, low collisionality pedestals and/or during the application of RMPs, as discussed in Section 4.3.2. The ratio of $L_\parallel$ to $L_{ELM}$ does not affect the total number of W atoms sputtered per ELM in the FSRM calculations, however; it only affects the peak value and characteristic decay rate of the ELMy W gross erosion rate. The relative contributions to this erosion rate from different impacting main ion and impurity species are also not affected by the chosen value of $L_{\parallel}$. In addition, as discussed in Section 2, the free-streaming framework remains valid as long as $L_{\parallel}/L_{ELM}$ > 5 [15].

The relevant parameters for the ITER reference magnetic equilibrium are described in [5, 46]. The electron temperature at the outer divertor target is set to 5 eV. A full-field ($B_T$ = 5.3 T) and half-field ($B_T$ = 2.65 T) case are investigated, with $n_{e,ped}$ and $T_{e,ped}$ taken from Table 2 in [3]. For the half-field case, each component of the magnetic field scaled down by a factor of 0.5, preserving the pitch angle of the magnetic field lines between both cases. No toroidal shaping of the divertor tiles is incorporated, as the emphasis is on general trends, but could be incorporated in a straightforward manner by changing $\theta_{div}$ from the value in the reference equilibrium.

In the FSRM-ITER calculations, the main ions in the pedestal are assumed to be both D and T in a 1:1 ratio. A 0.5% $Be^{4+}$ concentration at the pedestal top is assumed, similar to the values measured on JET-ILW [47]. Both a 1% $N^{7+}$ and 1% $Ne^{10+}$ pedestal impurity are considered to account for the expected level of impurity seeding in ITER [48]. In the full-field Q=10 case, a 2% $He^{2+}$ impurity is also included [8] to account for fusion alphas. It is not necessary to incorporate a recycling/re-deposition model for light impurities because the low divertor target temperature in ITER results in ion impact energies below the threshold for physical sputtering from any recycling ions. Heavy impurities in the pedestal and divertor are not included; i.e., the effect of W→W self-sputtering is neglected, which is justified in Section 2. The effect of free-streaming W ions from the pedestal top in ITER may be warranted in future investigation.

The calculated time evolution of the W erosion rate during unmitigated ELMs at the ITER outer divertor target is shown in Figure 14, with the contributions from D, T, He, Be, and N/Ne overlaid. The peak W sputtering rate caused by each species occurs at a slightly different time because the free-streaming ion velocity is inversely proportional to atomic mass, but the time delay between the peaks is small (~0.1 ms). The FSRM predicts gross W erosion during ELMs in ITER will be dominated by the main ions and seeded impurity, with only a small contribution from free-streaming $Be^{4+}$ and $He^{2+}$ ions. The contribution from free-streaming $T^{+}$ is approximately 2x larger than free-streaming $D^{+}$ due to the higher physical sputtering yield of T on W. As discussed above, there is zero contribution to W sputtering during ELMs from any recycling light ions. This result implies that extrapolations of the W divertor impurity source during ELMs in ITER are relatively insensitive to intrinsic impurities (Be or He), but *could be quite sensitive to both the species and injection rate of the seeded impurity*. For the full-field N case, the seeded impurity causes 25% of the total W erosion during ELMs, and seeded Ne is responsible 35% of the W erosion. Any change in the seeded impurity level in the pedestal will thus lead to a fairly substantial modification of the overall W sputtering rate during ELMs in the ITER divertor.

As expected, the peak and integrated W erosion during the full-field cases is substantially larger than the half-field case. However, despite a 4x increase in pedestal pressure, only about a 2x increase in total W erosion occurs because these high values of $T_{e,ped}$ correspond to the plateau regime where increasing impact energy of light ions does not materially affect the W physical sputtering yield. The total W erosion per ELM, per unit area, for each ITER case is plotted with the DIII-D results in Figure 10. The FSRM predicts that approximately 10x more W atoms will be sputtered per $m^2$ per ELM in the ITER divertor compared to DIII-D. While this result may initially seem alarming, it must be put into context of the additional physics that could regulate the W core impurity concentration in ITER.  A simple, 0D model for the core W concentration is given by

$$c_{W,core} = \frac{\Phi_W \cdot A_{div} \cdot f_{ELM} \cdot P_{leak} \cdot \tau_W}{n_{core} V_{core}} \tag{1}$$

where $\Phi_W$ is the total W atoms per area eroded per ELM, $f_{ELM}$ is the ELM frequency, $A_{div}$ is the divertor area, $P_{leak}$ is the W "leakage probability," i.e., the probability for an eroded W atom to contaminate the core, and $\tau_W$ is the effective W particle confinement time. This expression is normalized to the product of the core electron density, $n_{core}$, and the core volume, $V_{core}$. Interestingly, although $\Phi_W$ is predicted to be a factor of ~10 larger in ITER, the ratio $\Phi_W A_{div}/n_{core}V_{core}$ is nearly identical for both DIII-D and ITER. This is because core plasma volume increases as $R^3$ while the divertor plasma-wetted area only scales linearly with $R$. Therefore, for fixed values of $f_{ELM}$, $P_{leak}$, and $\tau_W$, the magnitude of W gross erosion during ELMs *normalized to the core particle content* may be very similar in ITER and current devices. This result is encouraging because it suggests it is appropriate to study ITER-relevant W sourcing and core contamination physics in current-generation tokamaks. The caveat, of course, is that ITER will operate in a regime of very low electron pedestal collisionality, $\nu^*_{e,ped}$ ~ 0.05-0.1, which is not the case for most of the discharges in the DIII-D database (or in general on most present devices).

The additional caveat is that high-power, full-field discharges in ITER must be operated in a regime of mitigated ELMs, i.e., low values of $\Delta W_{MHD}$ [2]. No first-principles model exists to predict the ELM size for ITER. ELM size has been extrapolated empirically based on the measured inverse relation with pedestal collisionality [49, 50], but this data has quite a large scatter and likely other physics effects are also important [16]. For a given value of $P_{heat}$, ELM size and frequency are widely observed to scale inversely, such that low ELM size implies high ELM frequency [43]. In Figure 15, the experimental measurements of the total intra-ELM W gross erosion rate per second, $\Phi_W \cdot f_{ELM}$, are plotted against ELM frequency from the DIII-D database. A strong correlation exists between the intra-ELM W sputtering source and ELM frequency, even at constant $P_{heat}$. For low-power (3-4 MW) discharges, the total ELMy W erosion source increases nearly linearly with $f_{ELM}$, indicating that $\Phi_W$ remains approximately constant even as ELM frequency increases. Interestingly, however, at higher heating power (5-7 MW) the W sputtering source pleateaus and may even 'roll-over' at sufficiently high ELM frequency. This is consistent with the previously discussed JET-ILW results [35] and suggests that ELM mitigation scenarios on ITER may be effective at reducing the overall intra-ELM W gross erosion if sufficiently high ELM frequencies can be achieved. In conjunction with the aforementioned high-Z impurity flushing effect often observed at high $f_{ELM}$ [36, 37, 35], the implications are promising for a low high-Z core contamination level in the ITER plasma core.

Equation (1) also implies that absolute predictions of the W core contamination level in ITER will require models for $P_{leak}$ and $\tau_W$. Investigation of such models is outside the scope of this paper, but it is noted that physics-based interpretive models exist for $P_{leak}$ and recent work is being performed to validate these models against DIII-D data [27, 51]. Uncertainties remain on a number of crucial physics parameters, such as the impact parallel flow and the effect of diffusive vs. convective cross-field impurity transport. In the pedestal region, high-Z confinement is generally modeled by inward neoclassical pinch forces between ELMs and strong anomalous outward transport during ELMs, suggesting a strong inverse dependence of the effective W confinement time on ELM frequency [36, 37]. Models for 'deep-core' W transport have recently become more sophisticated, e.g., [52], and include the physics of neoclassical convection, centrifugal asymmetries, turbulent transport, and MHD effects. Understanding the relative importance of these effects and the implications for ITER remains high-priority work in the field.

Finally, the relative contribution of ELMs to the overall W (intra-ELM + inter-ELM) gross erosion for each discharge in the DIII-D database is plotted in Figure 16. Shots from the gas puffing scan (166022, 26, 27) are excluded as the inter-ELM signal was too noisy for a reliable measurement. The ELMy contribution to the overall W gross erosion generally increases with ELM frequency, similar to what has been observed on JET-ILW [35]. A stronger correlation is observed with $f_{ELM}$ than with the overall W gross erosion source. Some evidence of a 'roll-over' exists in the DIII-D data for $f_{ELM}$ > 50 Hz, but is nearly within the error bars of the measurements. In all the attached DIII-D regimes studied in this work, the overall W erosion is mainly caused by physical sputtering during the inter-ELM phase. This is in contrast to the JET-ILW experience where the intra-ELM sputtering typically dominates [35, 53]. This difference is due to the higher physical sputtering yield of C→W relative to Be→W at the low impact energies (50-200 eV) characteristic of the attached inter-ELM phase and higher impurity concentrations present in the DIII-D divertor (1-2%) relative to the JET-ILW divertor (~0.5%). Detailed analysis of the spatial profile of the W sputtering in the DIII-D inter-ELM phase has previously been conducted and shown to be consistent with ERO+OEDGE modeling [20]. Overall, these results further reinforce the conclusion that ELMs will dominate the gross W erosion source in ITER, particularly close to the OSP.

## 6. Conclusions

The free-streaming plus recycling model (FSRM) for intra-ELM W gross erosion was described and tested against an extensive array of DIII-D experimental scenarios. A summary of the discharges analyzed and discussed in this work (i.e., those included in Figure 10) is provided in Table 1. The FSRM framework reveals that W sputtering during ELMs is dependent almost entirely on the density and temperature of the main ions and impurities at the top of the pedestal. The dependence on divertor conditions is relatively weak, entering only through the divertor temperature which determines the sheath potential and thus regulates the physical sputtering yield of recycling ions. The contribution of recycling ions tends to be relatively weak, however, for high $T_{e,epd}$, low $T_{e,div}$ scenarios of relevance to ITER and future devices. The FSRM was benchmarked against a DIII-D database of W sputtering measurements and found to agree well with the data at a variety of fueling levels, pedestal conditions, injected powers, and even ELM-mitigated scenarios, as long as the W sputtering measurement was conducted sufficiently close to the OSP. The FSRM may somewhat over-predict the W sputtering rate in some cases due to additional C deposits on the W-coated tiles in DIII-D, but no systematic dependence of the model/measurement discrepancies on the pedestal C impurity content was discovered.

This model was also tested in ELM-mitigated regimes achieved via pellet pacing and RMPs, techniques relevant for ITER. It was found that ELM mitigation via pellet pacing reduces W sputtering during ELMs by suppressing C impurity build-up in the pedestal and decreased the divertor target electron temperature, resulting in less physical sputtering of W caused by free-streaming $C^{6+}$ ions and recycling $C^{2+}$ ions, respectively. In contrast, no reduction of the W gross erosion rate was found during ELM mitigation via RMPs, contrary to the FSRM predictions. The RMP field also appeared to decrease the effective magnetic connection length between the center of the ELM filaments and the divertor targets. This result hints at important magnetic field effects occurring during the ELM-mitigated RMP scenarios that are currently not included in the model, such as magnetic field lines connecting directly to the divertor target from deeper into the plasma core.

The FSRM was extended to make predictions of the W gross erosion rate during unmitigated ELMs in ITER. It was found that the ELMy W sputtering source in ITER will likely be dominated by free-streaming D and T main ions, but there may also be a substantial contribution from the free-streaming ions of the chosen seeded impurity (N or Ne). Intrinsic pedestal impurities (He and Be) are not expected to contribute strongly to the W sputtering rate during ELMs, and divertor recycling will play no role due to the very lower divertor target temperature. It was also demonstrated that the total W gross erosion during ELMs normalized to the core particle content will be similar on ITER to the values in current devices. This underlines the importance of validated models to predict ELM frequency, high-Z divertor leakage, and core W confinement times in ITER to develop confidence that W core contamination in ITER will be suppressed to acceptable levels.

| Shot | Time (ms) | $P_{heat}$ (MW) | $n_{e,ped}$ ($10^{19}$ $m^{-3}$) | $T_{e,ped}$ (eV) | $f_{ELM}$ (Hz) | $\Phi_{W,Meas}$ ($10^{16}$ $m^{-2}$) | $\Phi_{W,Model}$ ($10^{16}$ $m^{-2}$) | Notes |
|---|---|---|---|---|---|---|---|---|
| 166022 | 1800-2600 | 5.7 | 3.8 | 660 | 51 | 88.4 | 101.9 | No-Puff Reference |
| 166026 | 1800-2600 | 6 | 4.9 | 525 | 61 | 91.6 | 100.8 | 70 Torr-L Gas Puff |
| 166027 | 2000-2600 | 5.9 | 6.2 | 360 | 71 | 55.2 | 72.9 | 200 Torr-L Gas Puff |
| 167297 | 2100-3000 | 3.6 | 7.3 | 363 | 9 | 74.7 | 81.1 | |
| 167298 | 2100-3000 | 3.7 | 7 | 367 | 10 | 75.3 | 86.8 | |
| 167300 | 2100-3000 | 3.6 | 5.8 | 361 | 26 | 54.8 | 72.2 | |
| 167317 | 2900-3500 | 6.2 | 4.2 | 496 | 68 | 41.3 | 53.6 | |
| 167320 | 4400-4900 | 4.9 | 4.4 | 428 | 48 | 81 | 62.3 | |
| 167321 | 2800-3500 | 6.4 | 4.1 | 540 | 63 | 54.5 | 55.1 | |
| 167322 | 2100-2600 | 3.6 | 4.8 | 435 | 36 | 62.5 | 65 | |
| 167322 | 4100-5000 | 3.6 | 5.4 | 403 | 23 | 72.6 | 70.3 | |
| 167354 | 2200-4400 | 3.5 | 6.8 | 355 | 10 | 74.8 | 79.6 | No-Pellet Reference |
| 167356 | 2600-4800 | 3.5 | 7.2 | 326 | 20 | 34.1 | 49.3 | 60 Hz Pellet Pacing |
| 167358 | 2600-4400 | 3.5 | 6.9 | 315 | 20 | 49.9 | 52.9 | 40 Hz Pellet Pacing |
| 167556 | 1800-2400 | 6.2 | 3.7 | 845 | 48 | 127.2 | 193.2 | No-RMP Reference |
| 167556 | 2600-3600 | 6.3 | 3 | 809 | 45 | 140.2 | 174.7 | RMP, n=3, 5 kA |

*Table 1. Summary of the DIII-D discharges analyzed and discussed in this paper, including comparison of the measured cumulative W erosion fluence per ELM, $\Phi_{W,meas}$, to the prediction of the FSRM, $\Phi_{W,Model}$.*

It is noted that this work only seeks to understand and validate a model for the gross erosion of tungsten during ELMs. As discussed above, the prompt re-deposition fraction of W during ELMs is likely very close to unity. Thus the net erosion of W is expected to be significantly reduced relative to the gross erosion. Recent ERO modeling has calculated re-deposition fractions of W during ELMs up to 99% in the JET-ILW divertor [54], but this study used an ELM-average approach with no time dependence. Refined models for prompt and non-prompt re-deposition should be incorporated to study the time evolution of W net erosion during the ELM cycle.

The results of this work highlight the important role that reduced, analytic models can play in understanding the physics of ELM-induced plasma materials interactions and for making extrapolations to future devices. These models may eventually replace expensive, non-linear MHD simulations of the ELM dynamics to identify the key physics parameters governing the incoming and outgoing particle fluxes in the tokamak divertor. Future work should focus on refining and further validating these models to address important physics effects such as: the 2D (radial+toroidal) spatial distribution of the tungsten sputtering flux; other mitigated ELM regimes such as Type III or "grassy" ELMs; the impact of other hydrogenic isotopes (H,T) and other intrinsic/seeded impurities on the W erosion rate; and the contributions of W prompt re-deposition and self-sputtering on the overall erosion picture. The eventual goal of such efforts should be the production of reduced-but-robust physics models such as scaling laws that can easily be extrapolated and applied to ELM regimes relevant for next-step devices.

## Acknowledgements

This material is based upon work supported by the U.S. Department of Energy, Office of Science, Office of Fusion Energy Sciences, using the DIII-D National Fusion Facility, a DOE Office of Science user facility, under Awards DE-FC02-04ER54698[a], DE-AC05-00OR22725[b], DE-FG02-07ER54917[c], DE-SC0013911[d], DE-AC04- 94AL8500[e] and DE-AC05-06OR23100[f]. DIII-D data shown in this paper can be obtained in digital format by following the links at https://fusion.gat.com/global/D3D_DMP. **Disclaimer:** This report was prepared as an account of work sponsored by an agency of the United States Government. Neither the United States Government nor any agency thereof, nor any of their employees, makes any warranty, express or implied, or assumes any legal liability or responsibility for the accuracy, completeness, or usefulness of any information, apparatus, product, or process disclosed, or represents that its use would not infringe privately owned rights. Reference herein to any specific commercial product, process, or service by trade name, trademark, manufacturer, or otherwise does not necessarily constitute or imply its endorsement, recommendation, or favoring by the United States Government or any agency thereof. The views and opinions of authors expressed herein do not necessarily state or reflect those of the United States Government or any agency thereof.

## Figure and Table Captions

Figure 1. Conceptual framework of the FSRM with magnetic field lines mapped into a 2D plane. An initially Gaussian plasma filament detaches into the SOL, where particles rapidly stream to the divertor targets and undergo repeated recycling.

Figure 2. Physical sputtering yields of C→W and D→W calculated by SDTrim.SP as a function of ion impact angle for the characteristic impact energy of 1000 eV.

Figure 3. FSRM predictions of the time evolution of the (a) divertor density, (b) divertor ion flux, (c) divertor heat flux, and (d) tungsten gross erosion rate from each sputtering mechanism as a function of time relative to the ELM start for typical DIII-D H-mode conditions with an attached divertor.

Figure 4. Fractional contributions of each W erosion mechanism to the gross erosion rate of tungsten during ELMs, calculated with the FSRM, as a function of typical DIII-D divertor and pedestal temperatures.

Figure 5. (a) Overview of the diagnostics used to perform ELM-resolved measurements in these studies to validate the FSRM. Magnified diagrams of the lower divertor region are also shown for (b) W-DiMES experiments and (c) the Metal Rings Campaign.

Figure 6. Time traces from a typical H-mode discharge during the DIII-D Metal Rings Campaign, displaying (a) line-average electron density, (b) MHD stored energy, (c) the radial position of the OSP, and (d) the WI emission intensity measured by a filterscope chord viewing one of the W-coated tiles on the divertor shelf.

Figure 7. An example of the ELM coherent averaging procedure used in this work. First the calibrated WI emission intensity signals are plotted as a function of time relative to the ELM start. Then a coherent average is produced by binning each time trace in 0.1 ms increments and averaging all the signals together, increasing the signal-to-noise ratio.

Figure 8. Measured W gross erosion per unit area as a function of pedestal density for a series of DIII-D discharges where the pedestal and divertor conditions were modified via gas puffing. The predictions of the FSRM, indicating the contribution of each erosion mechanism to the W sputtering rate, are overlaid.

Figure 9. Measured W sputtering rate for three different regimes explored by varying ELM size and frequency via heating power and plasma shaping. The predictions of the FSRM are overlaid, showing reasonable agreement in all cases.

Figure 10. Measured values of total eroded W atoms per ELM, plotted against the calculated values from the FSRM using $R_{eff} = 0.96$. Extrapolations to the ITER half-field and full-field case (Section 5) with N seeding are also overlaid.

Figure 11. (a) Ratio of the FSRM predictions of total W sputtered per ELM event to the experimental measurements, plotted against radial distance between the WI filterscope chord and the OSP. (b) The same ratio plotted as a function of pedestal C impurity content.

Figure 12. W gross erosion rates during ELMs as a function of ELM time (a) before and (b) after the application of pellet pacing to mitigate ELM size. The predictions of the FSRM, with each W sputtering mechanism indicated, are overlaid.

Figure 13. W gross erosion rates during ELMs as a function of ELM time (a) before and (b) after the application of the RMP to mitigate ELM size. The predictions of the FSRM, with each W sputtering mechanism indicated, are overlaid.

Figure 14. Predicted W gross erosion rates from the ITER outer target during ELMs for (a) the half-field case with N seeding, (b) the full-field case with N seeding, and (c) the full-field case with Ne seeding.

Figure 15. DIII-D experimental measurements of the total number of W atoms eroded per $m^2$ s during ELMs as a function of ELM frequency. The heating power level in each discharge is indicated by different symbol colors.

Figure 16. DIII-D experimental measurements of the ratio of the intra-ELM W gross erosion to the total W gross erosion, as a function of ELM frequency. The heating power level in each discharge is indicated by different symbol colors.

Table 1. Summary of the DIII-D discharges analyzed and discussed in this paper, including comparison of the measured cumulative W erosion fluence per ELM, $\Phi_{W,meas}$, to the prediction of the FSRM, $\Phi_{W,Model}$.

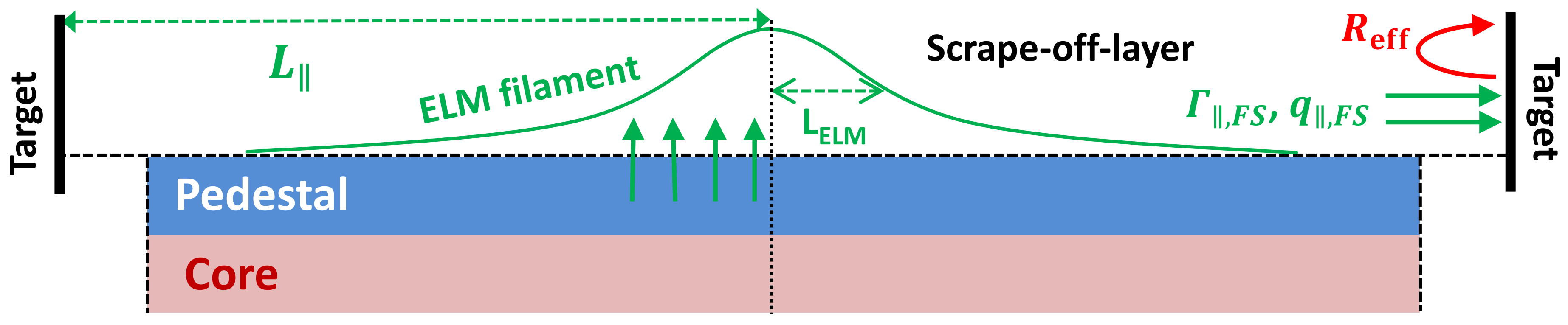

Target
$L_{\parallel}$
ELM filament
$L_{ELM}$
Scrape-off-layer
$R_{eff}$
$\Gamma_{\parallel,FS}$, $q_{\parallel,FS}$
Target
Pedestal
Core

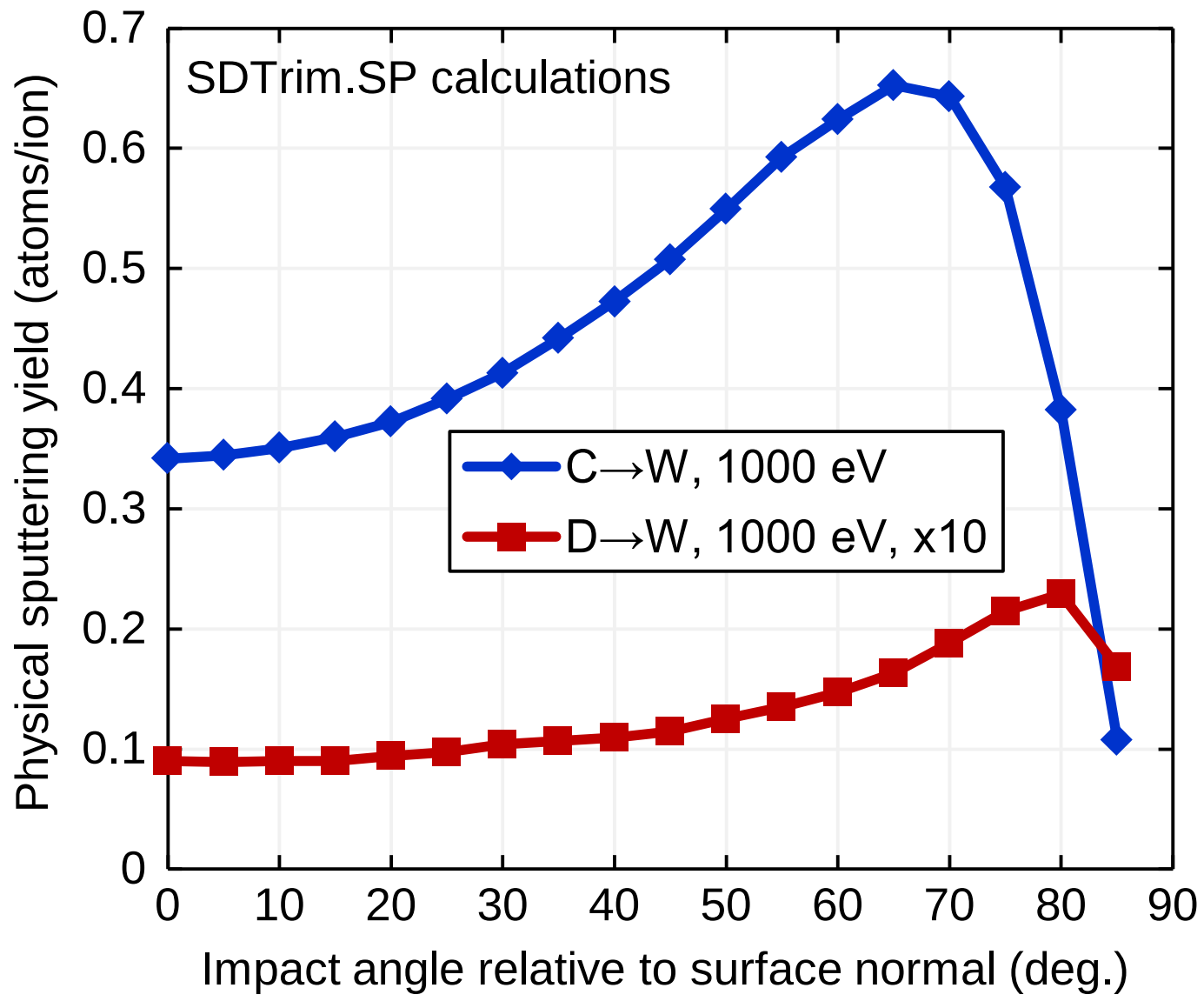

SDTrim.SP calculations
C→W, 1000 eV
D→W, 1000 eV, x10
Physical sputtering yield (atoms/ion)
Impact angle relative to surface normal (deg.)
0
0.1
0.2
0.3
0.4
0.5
0.6
0.7
0
10
20
30
40
50
60
70
80
90

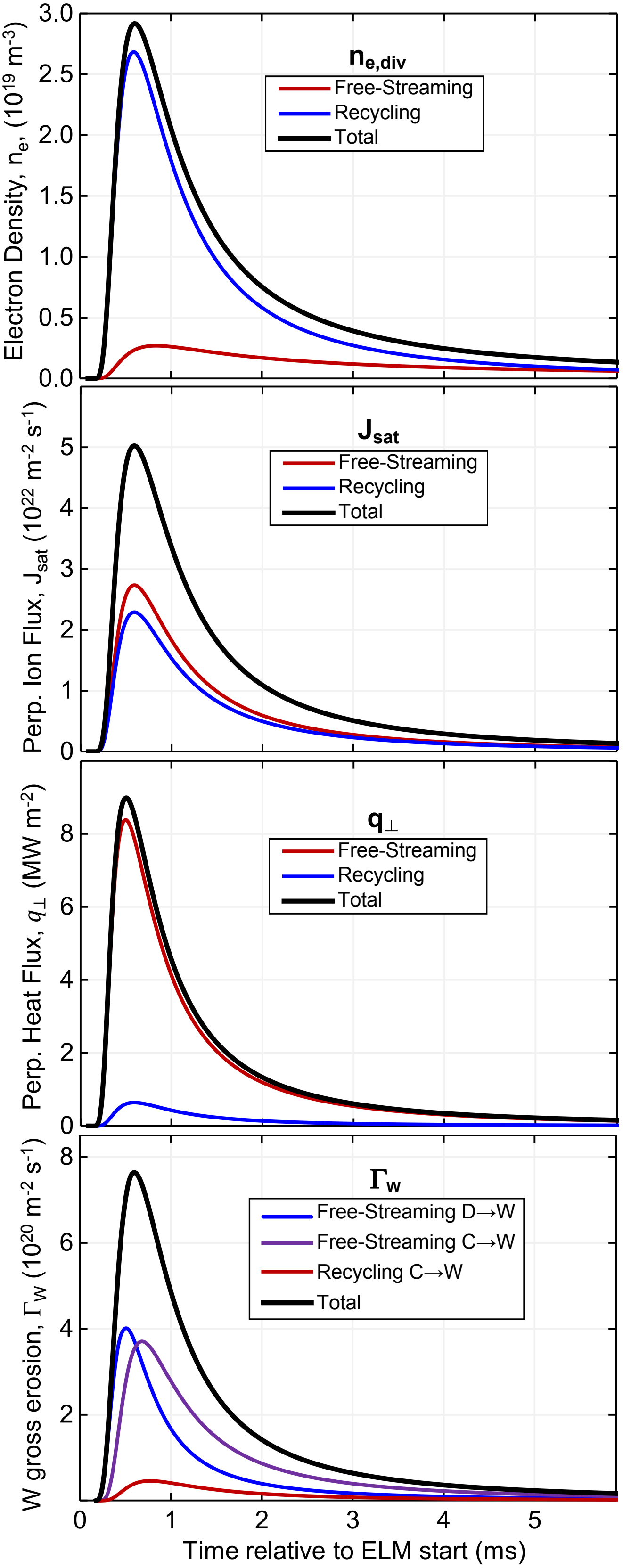

n_e,div
Free-Streaming
Recycling
Total
Electron Density, n_e, (10^19 m^-3)
3.0
2.5
2.0
1.5
1.0
0.5
0.0
J_sat
Free-Streaming
Recycling
Total
Perp. Ion Flux, J_sat (10^22 m^-2 s^-1)
5
4
3
2
1
0
q_⊥
Free-Streaming
Recycling
Total
Perp. Heat Flux, q_⊥ (MW m^-2)
8
6
4
2
0
Γ_W
Free-Streaming D→W
Free-Streaming C→W
Recycling C→W
Total
W gross erosion, Γ_W (10^20 m^-2 s^-1)
8
6
4
2
0
1
2
3
4
5
Time relative to ELM start (ms)

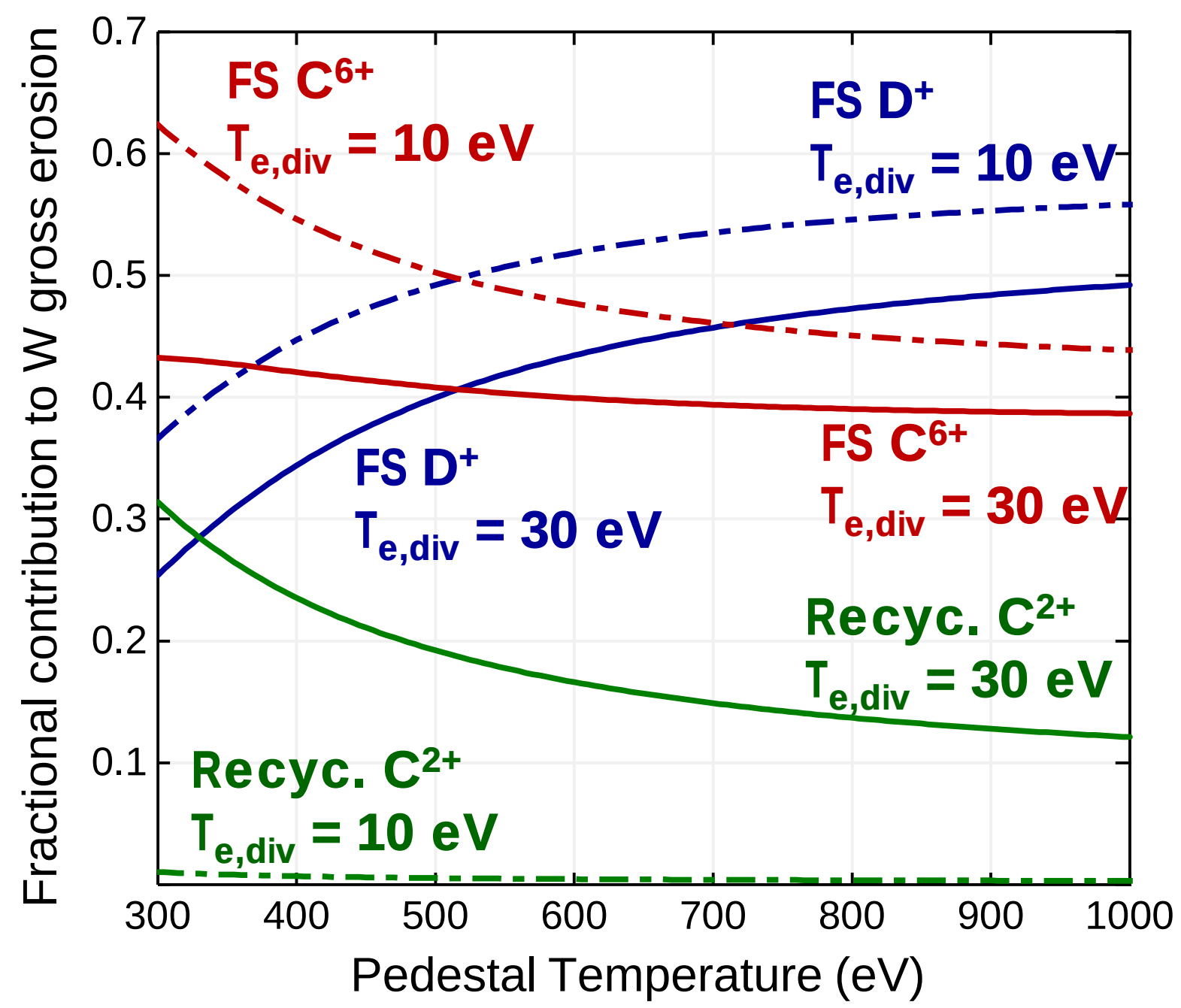

FS C6+ Te,div = 10 eV
FS D+ Te,div = 10 eV
FS D+ Te,div = 30 eV
FS C6+ Te,div = 30 eV
Recyc. C2+ Te,div = 30 eV
Recyc. C2+ Te,div = 10 eV
0.7
0.6
0.5
0.4
0.3
0.2
0.1
300
400
500
600
700
800
900
1000
Fractional contribution to W gross erosion
Pedestal Temperature (eV)

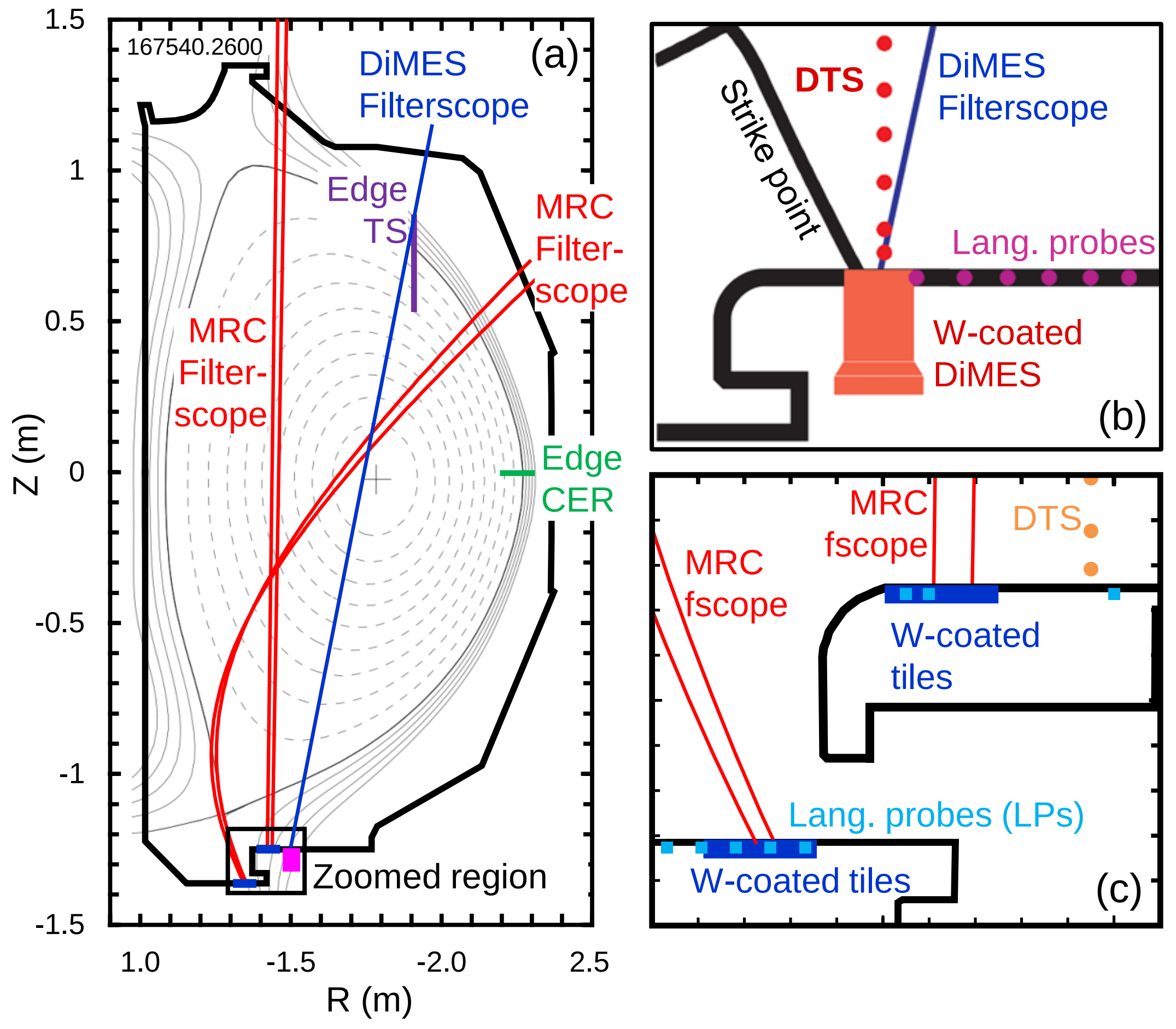

167540.2600
(a)
DiMES Filterscope
Edge TS
MRC Filter-scope
MRC Filter-scope
Edge CER
Zoomed region
Z (m)
R (m)
1.5
1
0.5
0
-0.5
-1
-1.5
1.0
-1.5
-2.0
2.5
DTS
DiMES Filterscope
Strike point
Lang. probes
W-coated DiMES
(b)
MRC fscope
DTS
MRC fscope
W-coated tiles
Lang. probes (LPs)
W-coated tiles
(c)

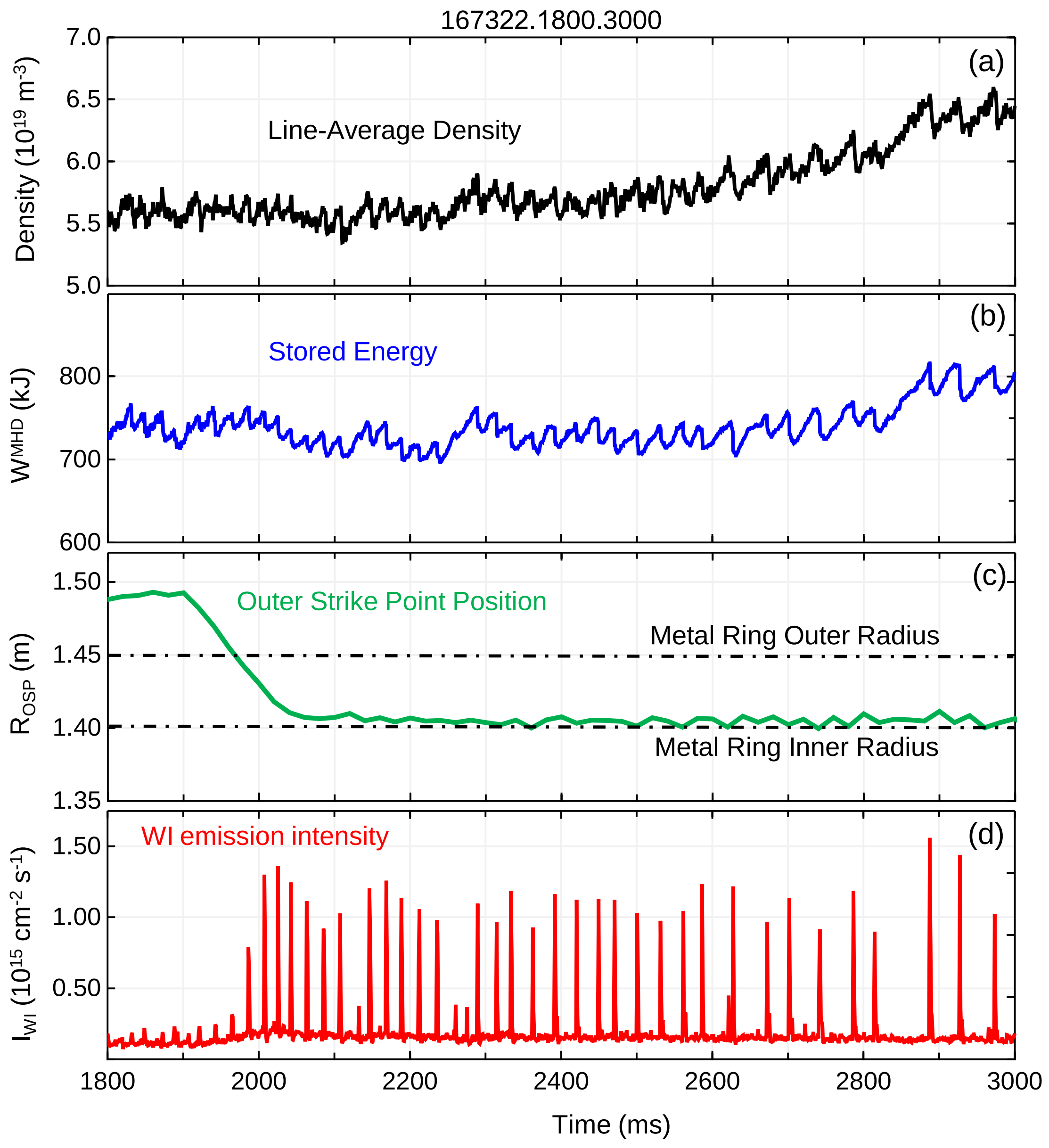
167322.1800.3000
(a)
Line-Average Density
Density ($10^{19}$ m$^{-3}$)
7.0
6.5
6.0
5.5
5.0
(b)
Stored Energy
$W^{MHD}$ (kJ)
800
700
600
(c)
Outer Strike Point Position
Metal Ring Outer Radius
Metal Ring Inner Radius
$R_{OSP}$ (m)
1.50
1.45
1.40
1.35
(d)
WI emission intensity
$I_{WI}$ ($10^{15}$ cm$^{-2}$ s$^{-1}$)
1.50
1.00
0.50
1800
2000
2200
2400
2600
2800
3000
Time (ms)

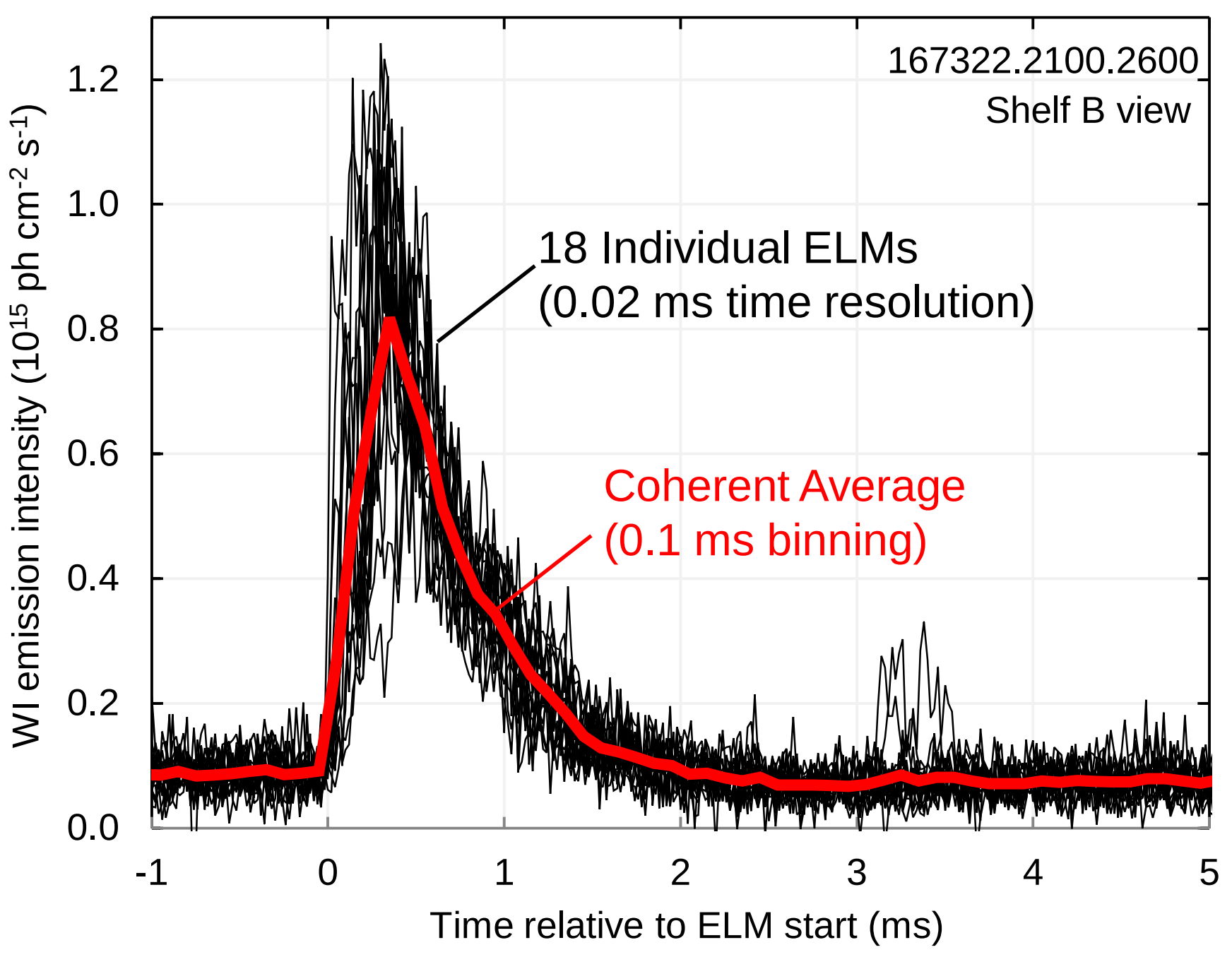

167322.2100.2600
Shelf B view
18 Individual ELMs
(0.02 ms time resolution)
Coherent Average
(0.1 ms binning)
WI emission intensity (10^15 ph cm^-2 s^-1)
Time relative to ELM start (ms)
1.2
1.0
0.8
0.6
0.4
0.2
0.0
-1
0
1
2
3
4
5

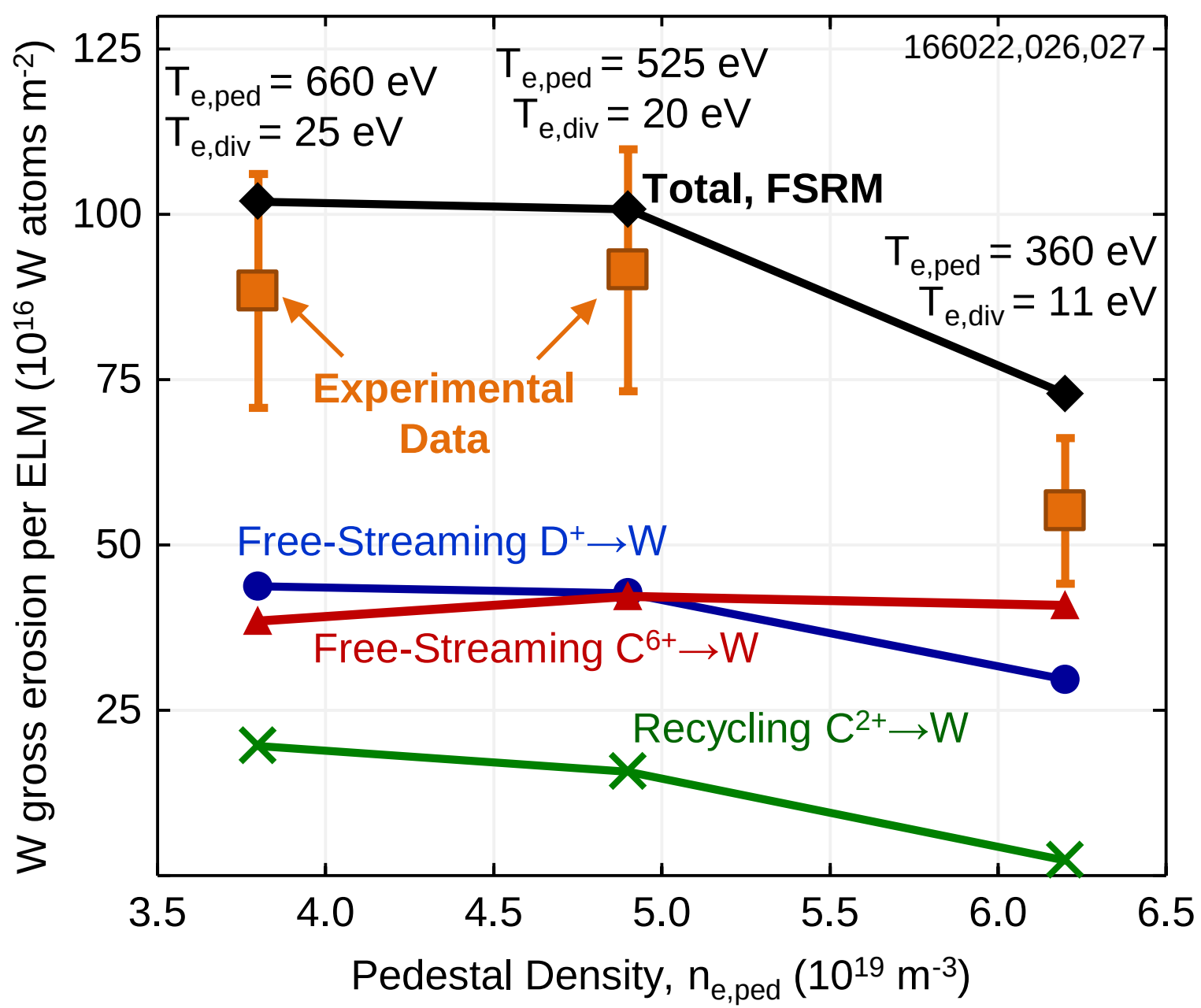

166022,026,027
T_e,ped = 660 eV
T_e,div = 25 eV
T_e,ped = 525 eV
T_e,div = 20 eV
Total, FSRM
T_e,ped = 360 eV
T_e,div = 11 eV
Experimental Data
Free-Streaming D+→W
Free-Streaming C6+→W
Recycling C2+→W
W gross erosion per ELM (10^16 W atoms m^-2)
Pedestal Density, n_e,ped (10^19 m^-3)
125
100
75
50
25
3.5
4.0
4.5
5.0
5.5
6.0
6.5

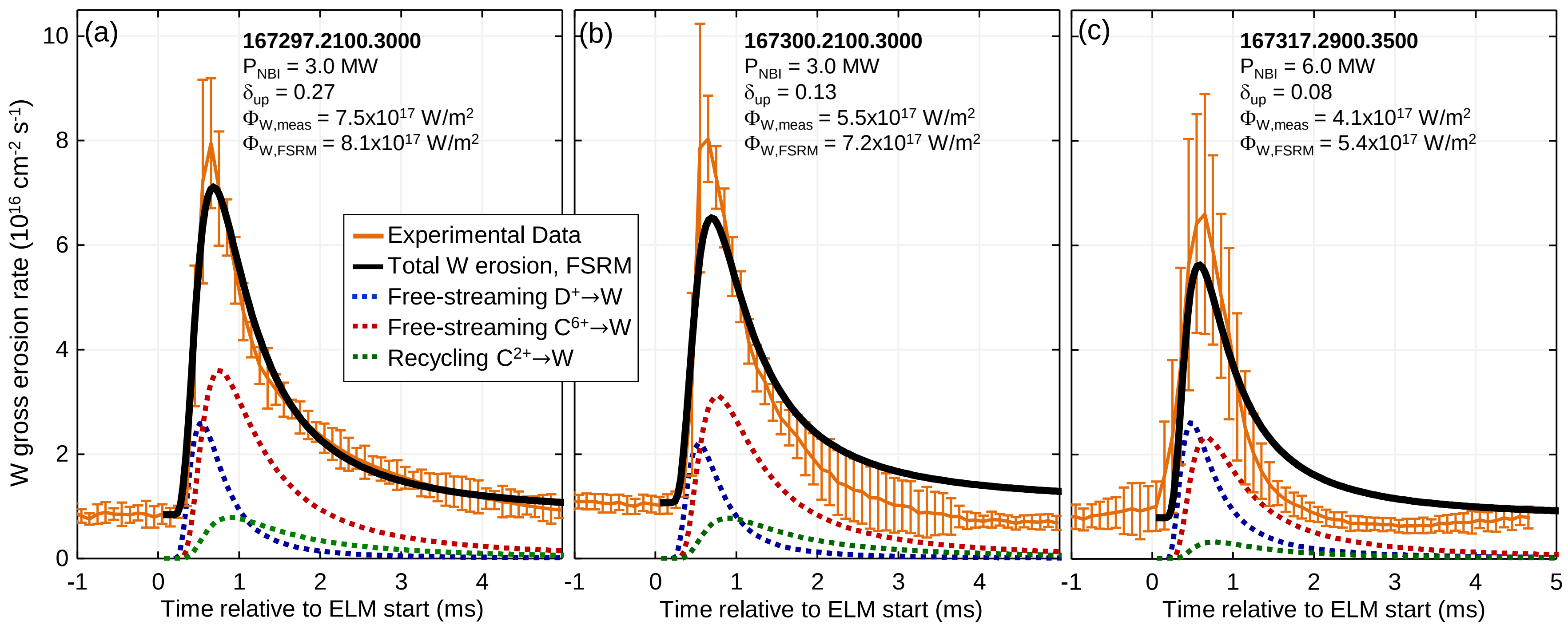

(a)
167297.2100.3000
$P_{NBI}$ = 3.0 MW
$\delta_{up}$ = 0.27
$\Phi_{W,meas}$ = 7.5x10$^{17}$ W/m$^2$
$\Phi_{W,FSRM}$ = 8.1x10$^{17}$ W/m$^2$
(b)
167300.2100.3000
$P_{NBI}$ = 3.0 MW
$\delta_{up}$ = 0.13
$\Phi_{W,meas}$ = 5.5x10$^{17}$ W/m$^2$
$\Phi_{W,FSRM}$ = 7.2x10$^{17}$ W/m$^2$
(c)
167317.2900.3500
$P_{NBI}$ = 6.0 MW
$\delta_{up}$ = 0.08
$\Phi_{W,meas}$ = 4.1x10$^{17}$ W/m$^2$
$\Phi_{W,FSRM}$ = 5.4x10$^{17}$ W/m$^2$
Experimental Data
Total W erosion, FSRM
Free-streaming $D^+$→W
Free-streaming $C^{6+}$→W
Recycling $C^{2+}$→W
W gross erosion rate (10$^{16}$ cm$^{-2}$ s$^{-1}$)
Time relative to ELM start (ms)
Time relative to ELM start (ms)
Time relative to ELM start (ms)

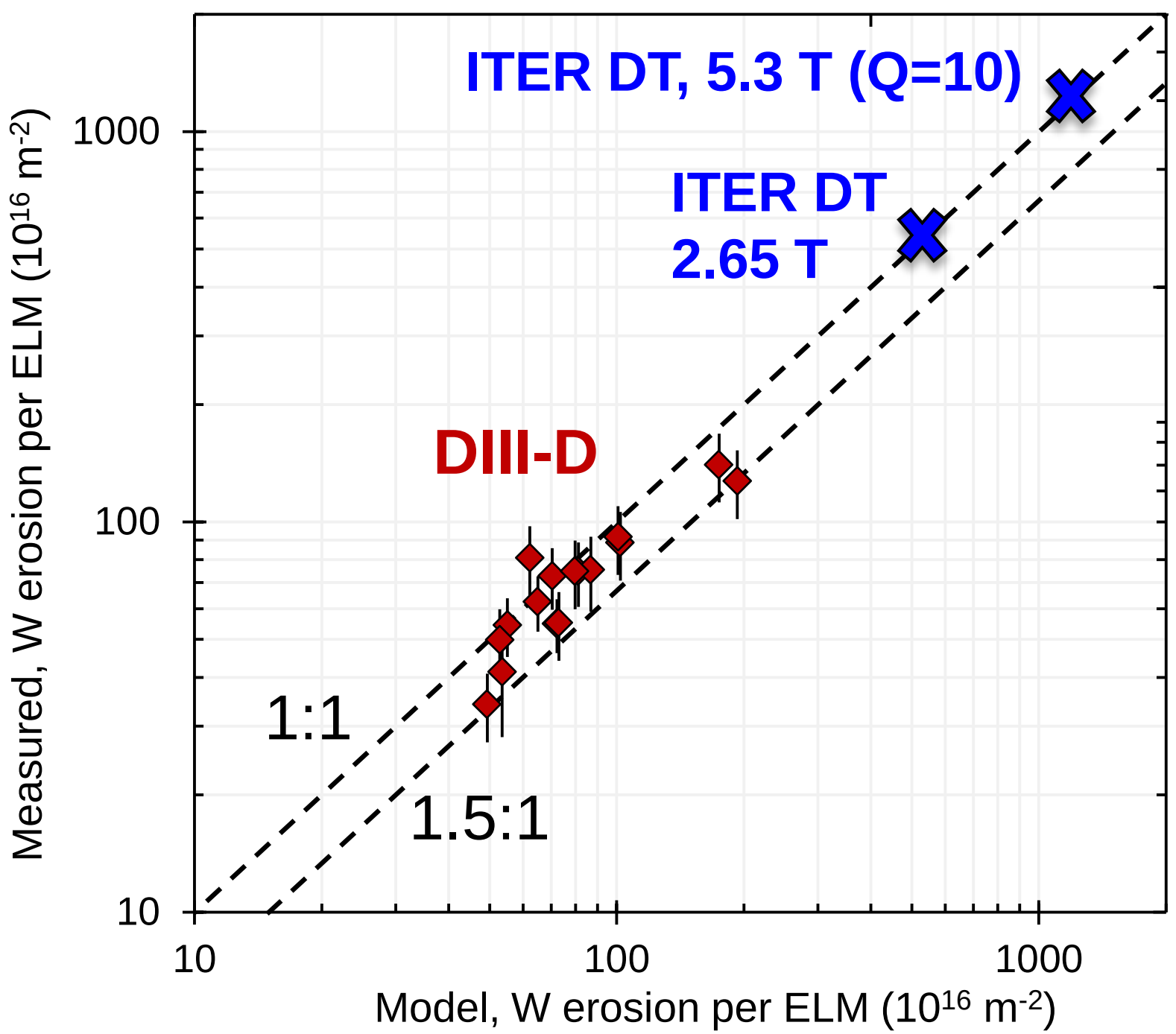

ITER DT, 5.3 T (Q=10)
ITER DT
2.65 T
DIII-D
1:1
1.5:1
1000
100
10
10
100
1000
Measured, W erosion per ELM ($10^{16}$ m$^{-2}$)
Model, W erosion per ELM ($10^{16}$ m$^{-2}$)

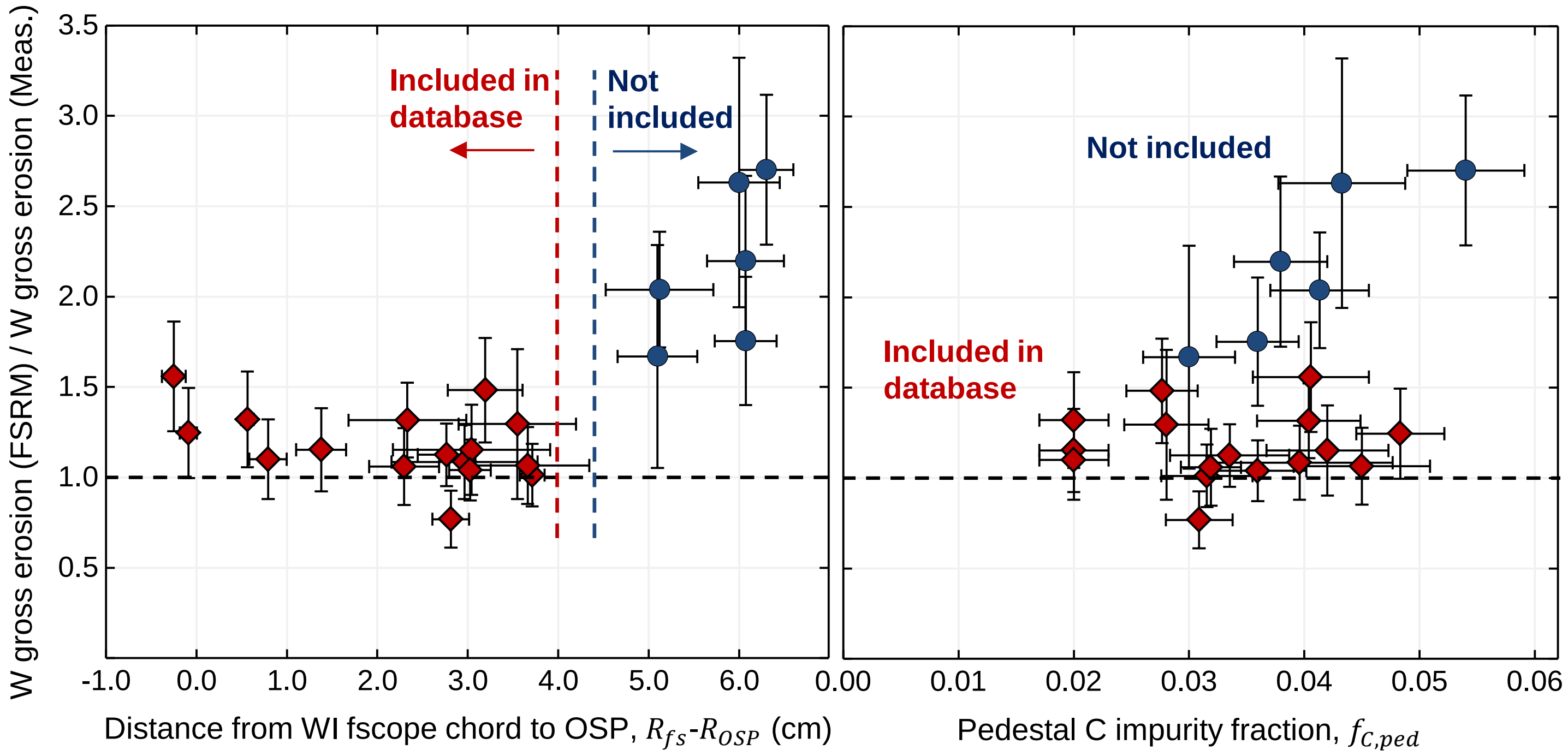

W gross erosion (FSRM) / W gross erosion (Meas.)
3.5
3.0
2.5
2.0
1.5
1.0
0.5
Included in database
Not included
-1.0
0.0
1.0
2.0
3.0
4.0
5.0
6.0
Distance from WI fscope chord to OSP, $R_{fs}$-$R_{OSP}$ (cm)
Not included
Included in database
0.00
0.01
0.02
0.03
0.04
0.05
0.06
Pedestal C impurity fraction, $f_{C,ped}$

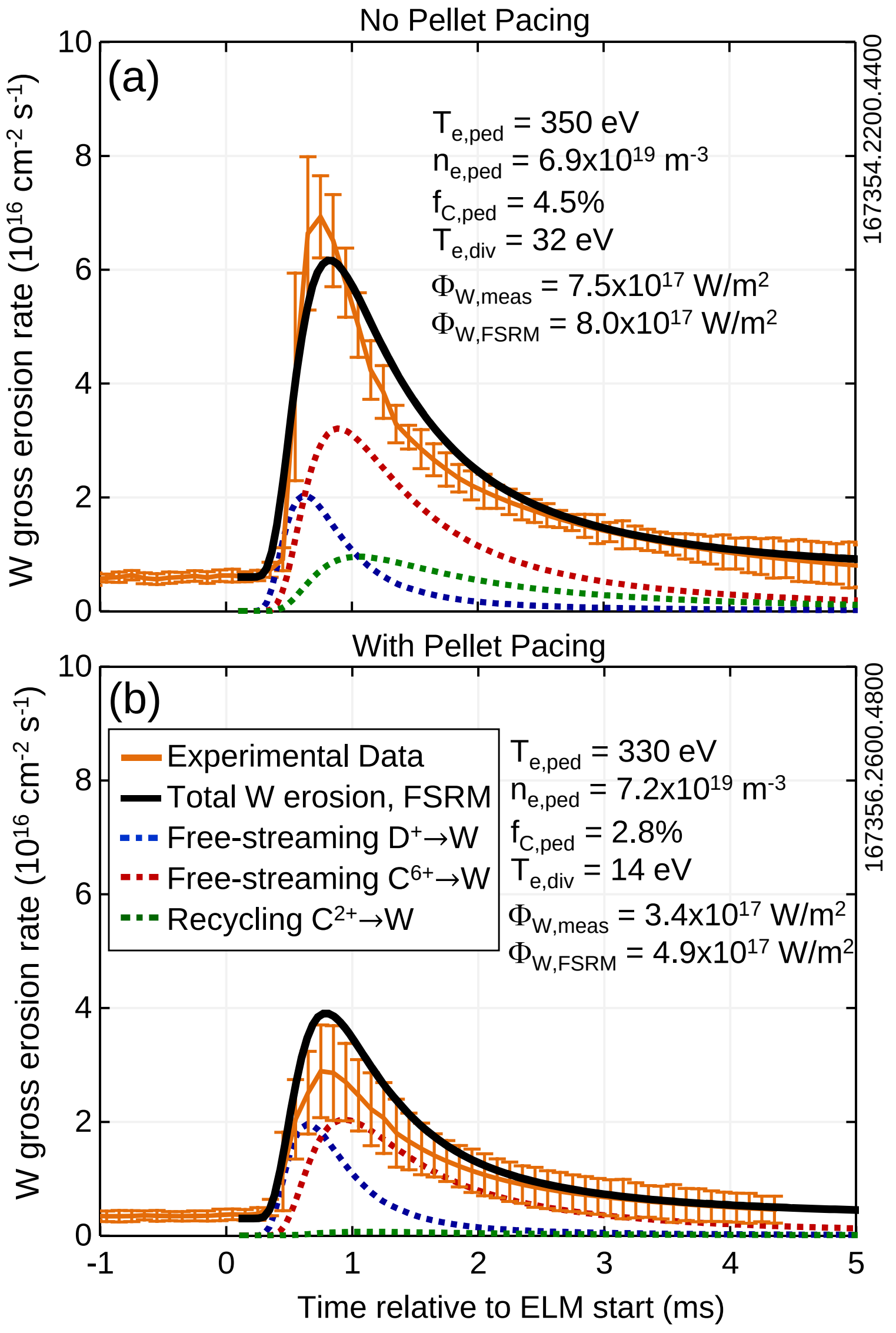
No Pellet Pacing
(a)
T_e,ped = 350 eV
n_e,ped = 6.9x10^19 m^-3
f_C,ped = 4.5%
T_e,div = 32 eV
Φ_W,meas = 7.5x10^17 W/m^2
Φ_W,FSRM = 8.0x10^17 W/m^2
167354.2200.4400
W gross erosion rate (10^16 cm^-2 s^-1)
With Pellet Pacing
(b)
Experimental Data
Total W erosion, FSRM
Free-streaming D+→W
Free-streaming C6+→W
Recycling C2+→W
T_e,ped = 330 eV
n_e,ped = 7.2x10^19 m^-3
f_C,ped = 2.8%
T_e,div = 14 eV
Φ_W,meas = 3.4x10^17 W/m^2
Φ_W,FSRM = 4.9x10^17 W/m^2
167356.2600.4800
0
2
4
6
8
10
-1
1
3
5
Time relative to ELM start (ms)

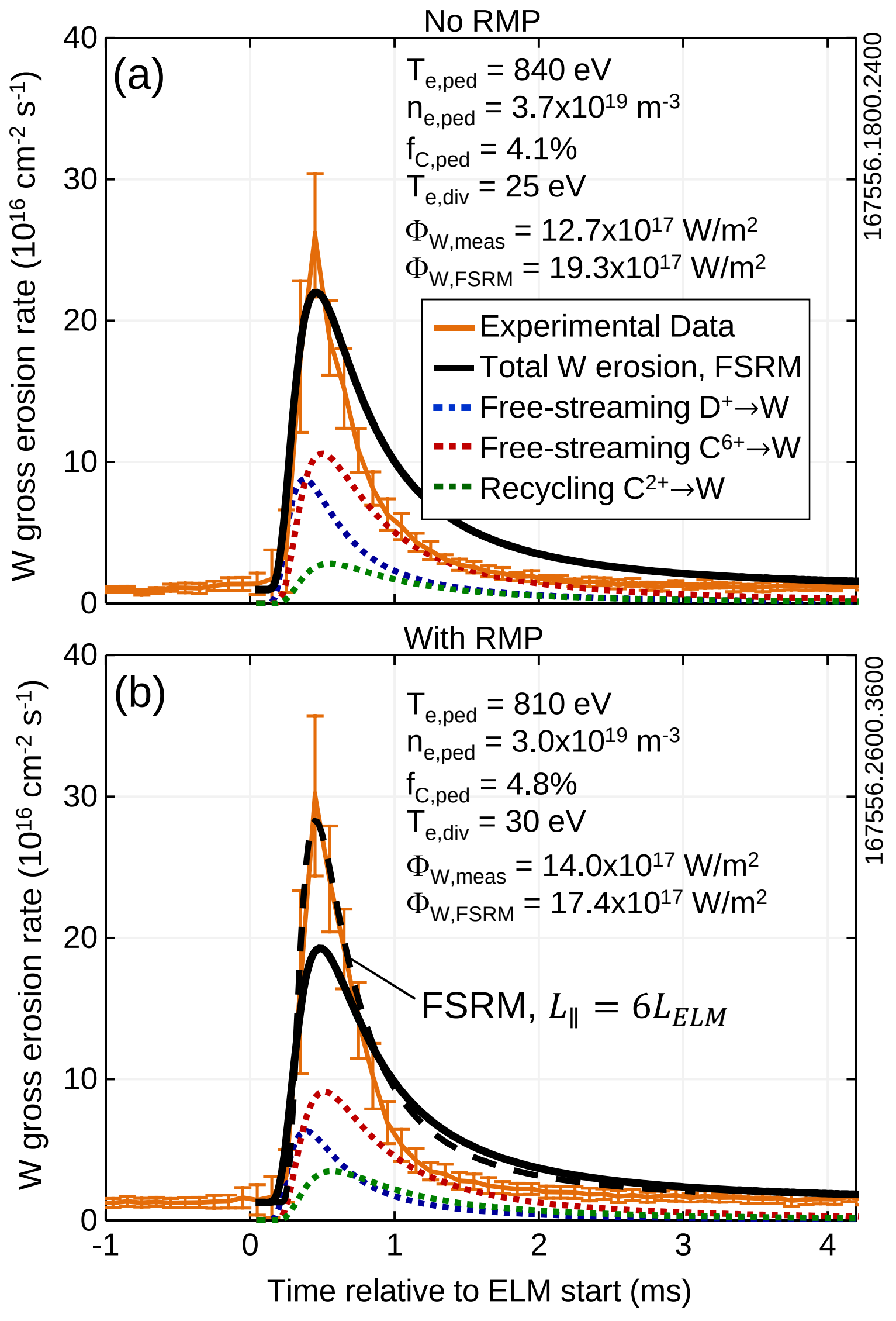

No RMP
(a)
T_e,ped = 840 eV
n_e,ped = 3.7x10^19 m^-3
f_C,ped = 4.1%
T_e,div = 25 eV
Φ_W,meas = 12.7x10^17 W/m^2
Φ_W,FSRM = 19.3x10^17 W/m^2
Experimental Data
Total W erosion, FSRM
Free-streaming D+→W
Free-streaming C6+→W
Recycling C2+→W
167556.1800.2400
With RMP
(b)
T_e,ped = 810 eV
n_e,ped = 3.0x10^19 m^-3
f_C,ped = 4.8%
T_e,div = 30 eV
Φ_W,meas = 14.0x10^17 W/m^2
Φ_W,FSRM = 17.4x10^17 W/m^2
FSRM, L_∥ = 6L_ELM
167556.2600.3600
W gross erosion rate (10^16 cm^-2 s^-1)
Time relative to ELM start (ms)
0
10
20
30
40
-1
0
1
2
3
4

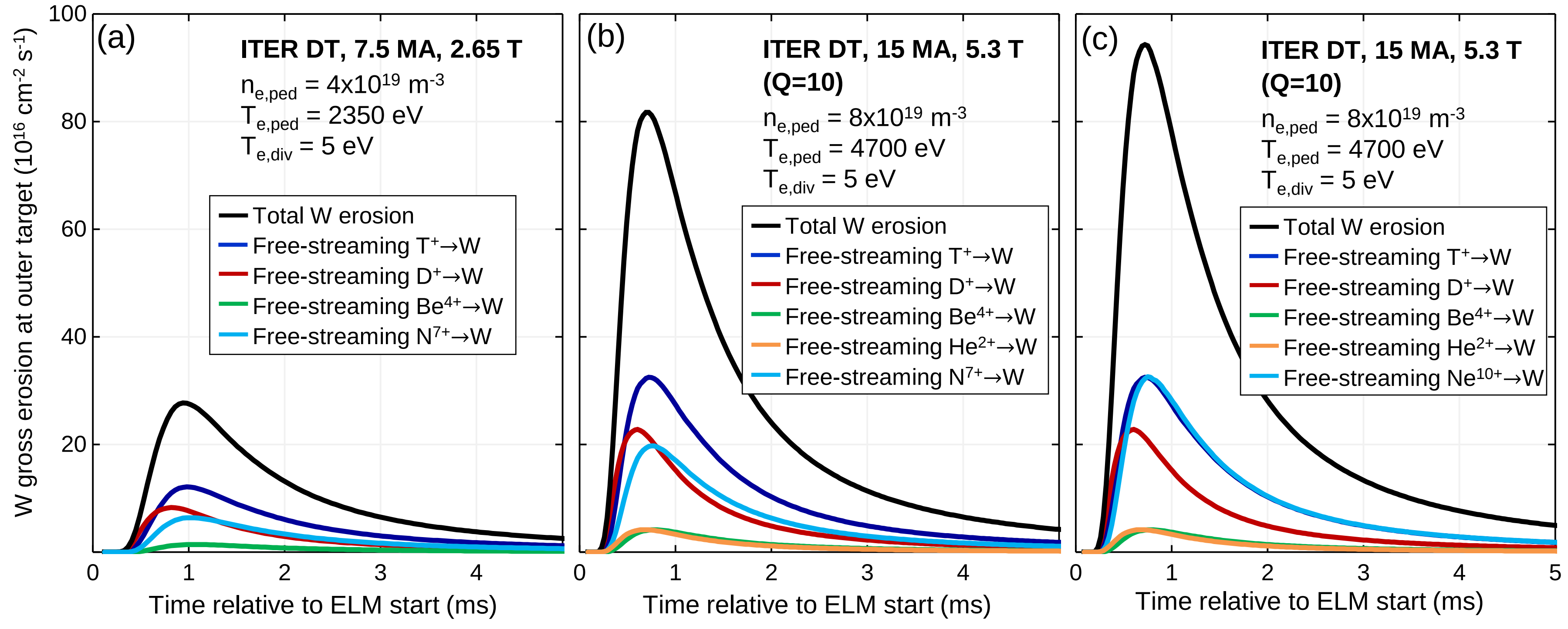

W gross erosion at outer target ($10^{16}$ cm$^{-2}$ s$^{-1}$)
Time relative to ELM start (ms)
(a)
ITER DT, 7.5 MA, 2.65 T
$n_{e,ped}$ = 4x10$^{19}$ m$^{-3}$
$T_{e,ped}$ = 2350 eV
$T_{e,div}$ = 5 eV
Total W erosion
Free-streaming $T^+$→W
Free-streaming $D^+$→W
Free-streaming $Be^{4+}$→W
Free-streaming $N^{7+}$→W
(b)
ITER DT, 15 MA, 5.3 T (Q=10)
$n_{e,ped}$ = 8x10$^{19}$ m$^{-3}$
$T_{e,ped}$ = 4700 eV
$T_{e,div}$ = 5 eV
Total W erosion
Free-streaming $T^+$→W
Free-streaming $D^+$→W
Free-streaming $Be^{4+}$→W
Free-streaming $He^{2+}$→W
Free-streaming $N^{7+}$→W
(c)
ITER DT, 15 MA, 5.3 T (Q=10)
$n_{e,ped}$ = 8x10$^{19}$ m$^{-3}$
$T_{e,ped}$ = 4700 eV
$T_{e,div}$ = 5 eV
Total W erosion
Free-streaming $T^+$→W
Free-streaming $D^+$→W
Free-streaming $Be^{4+}$→W
Free-streaming $He^{2+}$→W
Free-streaming $Ne^{10+}$→W

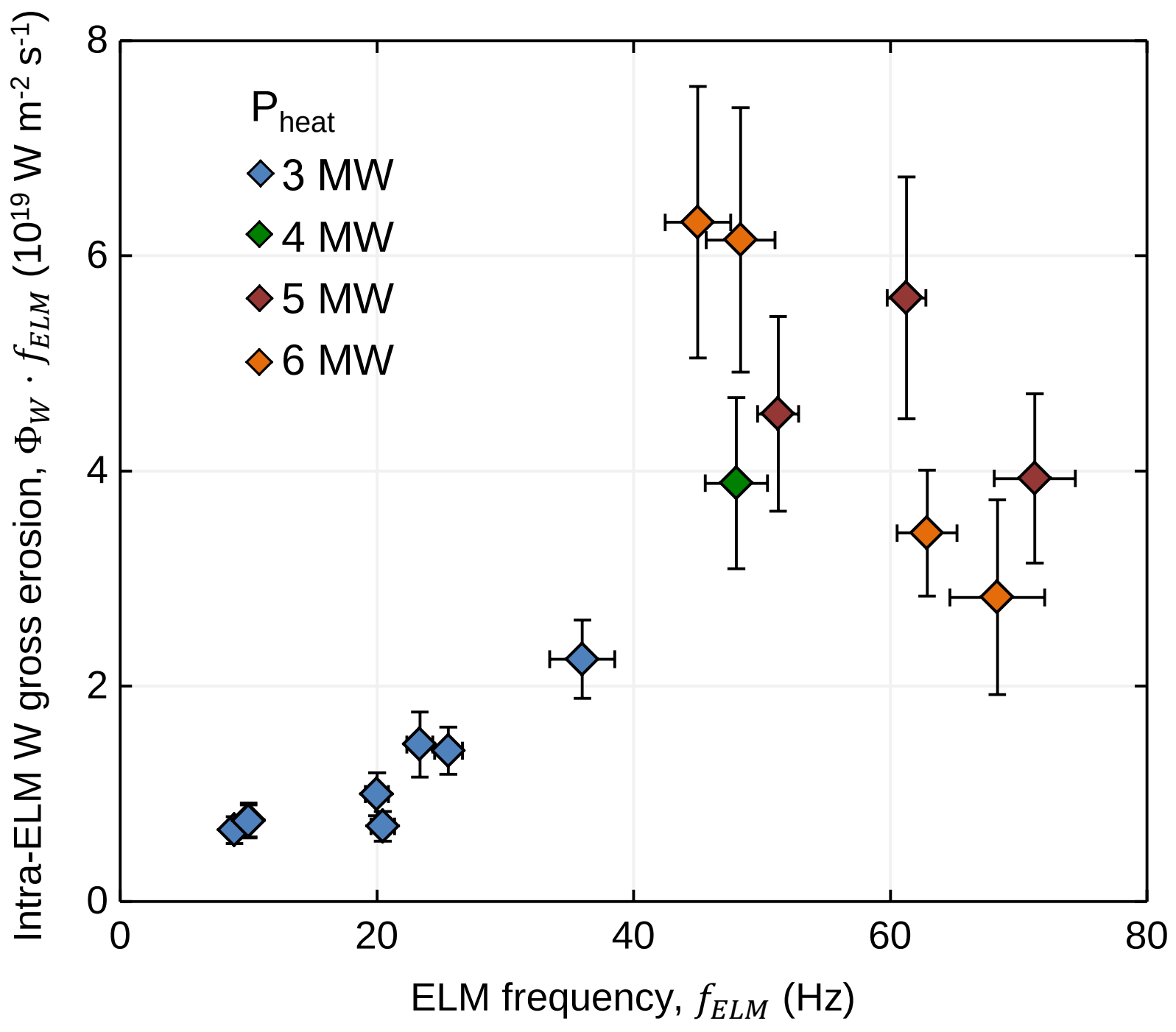

Intra-ELM W gross erosion, $\Phi_W \cdot f_{ELM}$ ($10^{19}$ W m$^{-2}$ s$^{-1}$)
ELM frequency, $f_{ELM}$ (Hz)
$P_{heat}$
3 MW
4 MW
5 MW
6 MW
0
2
4
6
8
0
20
40
60
80

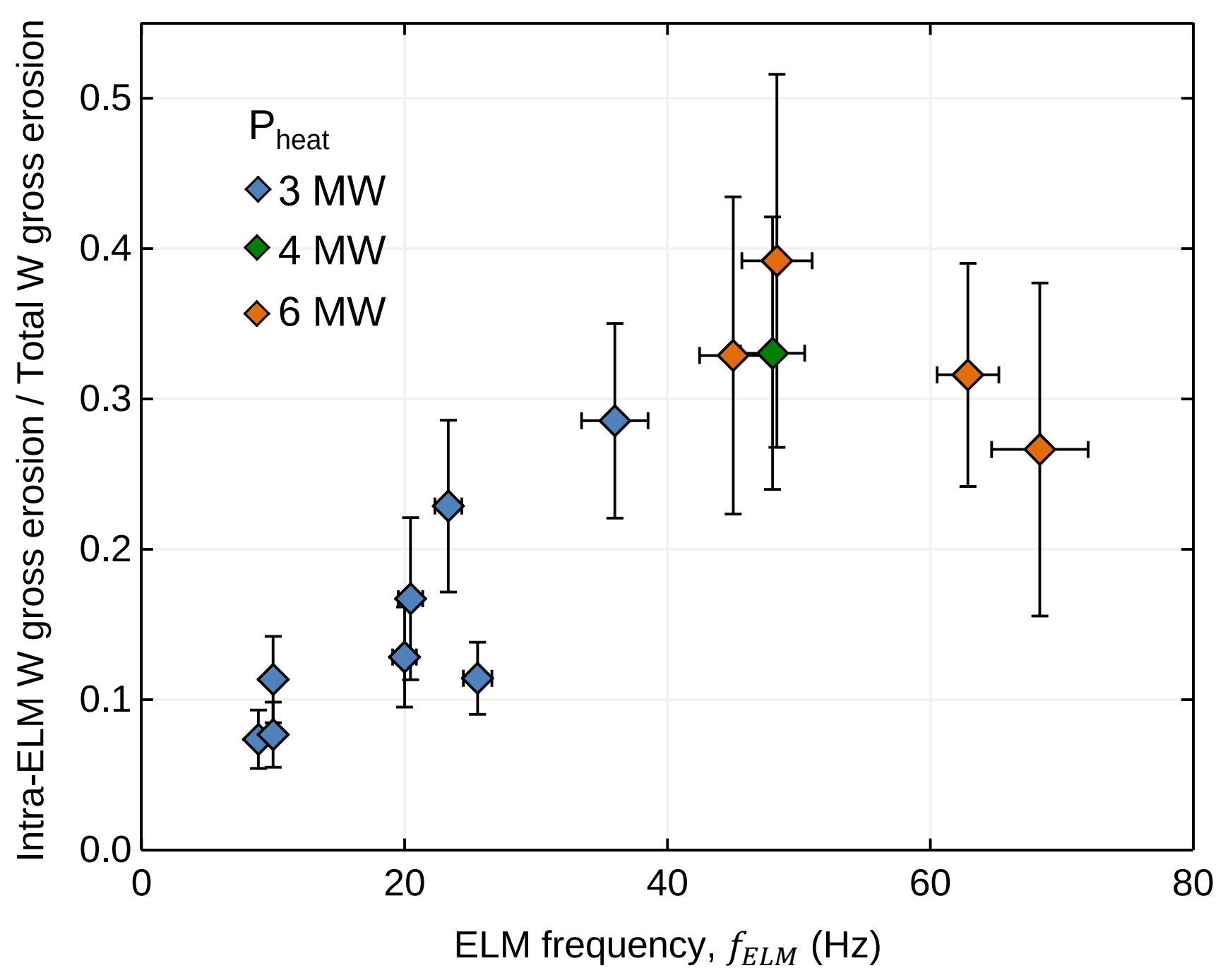

Intra-ELM W gross erosion / Total W gross erosion
0.0
0.1
0.2
0.3
0.4
0.5
0
20
40
60
80
ELM frequency, $f_{ELM}$ (Hz)
$P_{heat}$
3 MW
4 MW
6 MW